\documentclass[pss]{wiley2sp} 
\usepackage{amsmath}

\begin{document}

\title{Ferroelectricity from iron valence ordering in rare earth ferrites?}

\titlerunning{Rare earth ferrites }

\author{%
  Manuel Angst\textsuperscript{\Ast,\textsf{\bfseries 1},\textsf{\bfseries 2}}}

\authorrunning{M. Angst}

\mail{e-mail
  \textsf{m.angst@fz-juelich.de}, Phone:
  +49-2461-612479}

\institute{%
  \textsuperscript{1}\, Peter Gr{\"u}nberg Institut PGI and J{\"u}lich Centre for Neutron Science JCNS, JARA-FIT, Forschungszentrum J{\"u}lich GmbH, 52425
  J{\"u}lich, Germany\\
  \textsuperscript{2}\, Experimental Physics IVC, RWTH Aachen University, 52056 Aachen, Germany}

\received{XXXX, revised XXXX, accepted XXXX} 
\published{19 April 2013 - Phys.\ Status Solidi RRL, DOI:10.1002/pssr.201307103 (2013)} 


\abstract{%
\abstcol{%
}
{%
  The possibility of multiferroicity arising from charge ordering in LuFe$_2$O$_4$ and
  structurally related rare earth ferrites is reviewed. Recent experimental work
  on macroscopic indications of ferroelectricity and microscopic determination of
  coupled spin and charge order indicates that this scenario does not hold.
  Understanding the origin of the experimentally observed charge and spin order
  will require further theoretical work. Other aspects of recent research in these
  materials, such as geometrical frustration effects, possible electric-field-induced
  transitions, or orbital order are also briefly treated. }}

\titlefigure[height=3.1cm]{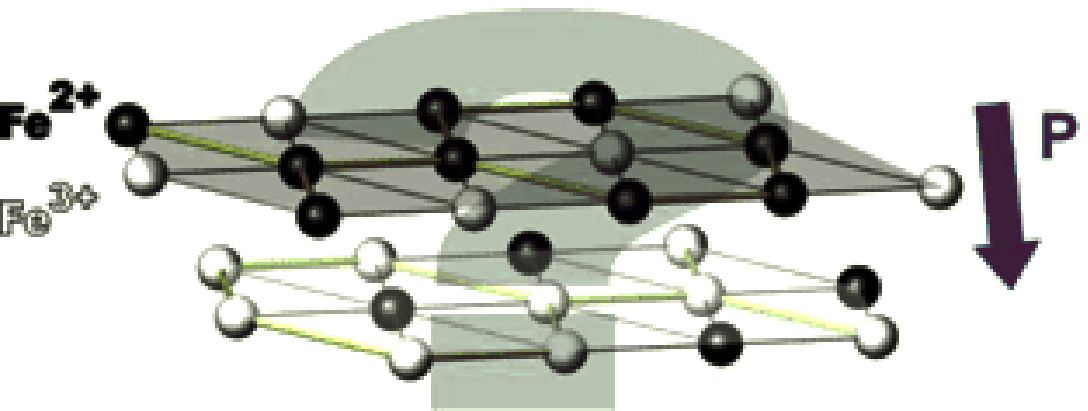}
\titlefigurecaption{%
  Bilayer subunit in $R$Fe$_2$O$_4$ compounds with proposed charge order rendering it polar. }

\maketitle   

\section{Introduction}

The interplay of magnetic and electric degrees of freedom in
correlated electron materials is one of the major topics of
contemporary condensed matter physics, with a rapidly rising
number of publications per year \cite{Fiebig09}. Materials
combining magnetism with ferroelectricity, termed
multiferroics, offer the prospect of a large magnetoelectric
coupling, which has a high potential for applications in future
information technology \cite{Bibes08}. Because the traditional
mechanism of ferroelectricity is incompatible with magnetism
\cite{Hill00}, much interest has focused on unconventional
routes to ferroelectricity (see \cite{Wang09} for a recent
extensive review). Particularly intriguing is the mechanism of
ferroelectricity originating from charge ordering (CO)
(reviewed in \cite{Brink08,Sun13}), the ordered arrangement of
different valence states of an ion, typically a transition
metal. Because any ferroelectric polarization is built from
electric dipole moments, i.e.\ non-centrosymmetric charge
distributions, any CO breaking inversion symmetry automatically
induces a polarization, which may be very large \cite{Brink08}.
The presence of different valence states of a transition metal
ion implies an active spin degree of freedom on the same ion,
and therefore a strong magnetoelectric coupling may be expected
as well. This mechanism of multiferroicity is thus very
attractive from the point of view of prospective applications.
However, while the concept is straight-forward, examples of
oxides where this mechanism is experimentally indicated to
occur are exceedingly rare and none is really well understood.

\begin{figure*}[t]%
  \sidecaption
\includegraphics*[width=.65\linewidth]{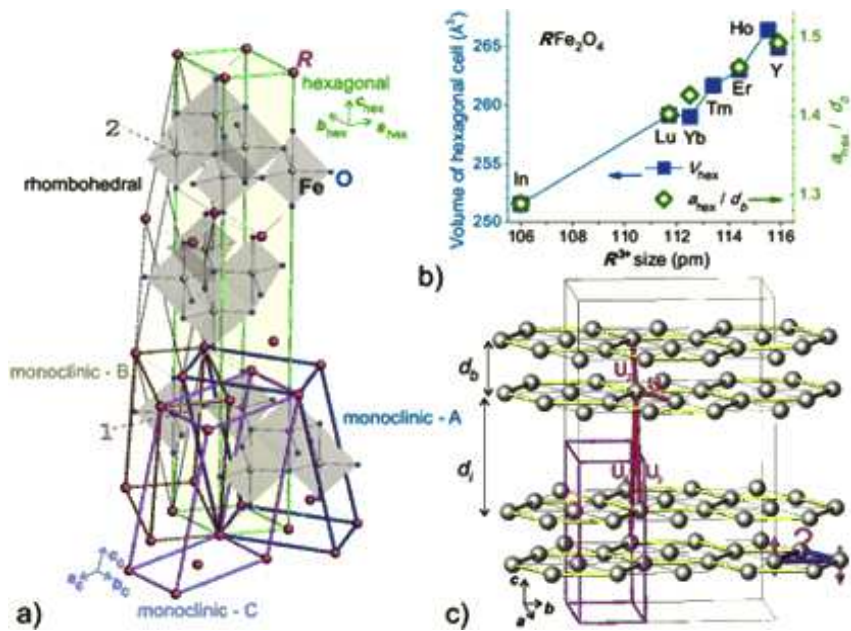}
\caption{%
  a) $R\overline{3}m$ Crystal structure of $R$Fe$_2$O$_4$ and description with different cells (after \cite{Xu10}). Apart from the primitive rhombohedral and the
$R$-centered hexagonal cells the lattice could also be
described by one of three $C$-centered monoclinic cells, which
are rotated by 120$^\circ$ with respect to each other. Cells
describing the charge and spin order in the three domains are
obtained from these by tripling the monoclinic $b$ and doubling
the monoclinic $c$ axes (omitted for clarity). b) $R^{3+}$ ion
size effect on cell volume and ratio of intralayer Fe$-$Fe
distance $a_{\rm hex}$ to bilayer thickness $d_b$ (c.f. c),
compiled from
\cite{Kato75,Isobe90,Matsumoto92b,Oka08,Yoshii06,Yoshii08b,Yoshii08}.
c) Iron bilayers with one of the monoclinic cells, and the
corresponding supercell. Blue triangle illustrates geometrical
frustration effects (see text). Red lines denote a minimal set
of interactions \cite{Yamada00,Harris10} necessary for 3D
charge order.} \label{compstrucfig}
\end{figure*}

The often cited (e.g.\ \cite{Wang09,Brink08,Sun13})
prototypical example material providing ``proof of principle''
for this mechanism is LuFe$_2$O$_4$, a charge- (and spin-)
frustrated system. In the following, I will review recent
research on this and isostructural materials, which may serve
also as an illuminating case study for (much less investigated)
other potential examples of CO-based ferroelectricity.
LuFe$_2$O$_4$ had been proposed in 2005 to be a multiferroic
due to Fe$^{2+/3+}$ CO \cite{Ikeda05}, based on both
macroscopic indications by dielectric spectroscopy and
pyroelectric current measurements, and on a plausible
microscopic model of a polar CO. This proposed
``ferroelectricity from CO'' is the main focus of the review,
in particular recent findings by structure refinement of a
non-polar CO \cite{DeGroot12b} and the suggestion of the
absence of polar order by dielectric spectroscopy
\cite{Niermann12,Ruff12}. Other aspects that have been of
recent focus in these materials, such as geometrical
frustration expressed in CO and magnetic order, strong
spin-charge coupling, or potential electric-field-induced phase
transitions, are treated as well. The main experimental
findings reviewed are summarized in Sec.\ \ref{sum}, which
concludes with a brief outlook on future research directions
both in this family of compounds and for ``ferroelectricity
from CO'' in general.

\section{The $R$Fe$_2$O$_4$ family}
\label{family}

Rare earth ferrites $R$Fe$_2$O$_4$ with $R$ a $3+$ ion without
partially occupied $d$ levels (Y, Ho, Er, Tm, Yb, Lu, or In),
known since the 1970s \cite{Kimizuka75}, crystallize in a
rhombohedral ($R\overline{3}m$) structure
\cite{Kato75,Isobe90,Matsumoto92b,Oka08,Yoshii06,Yoshii08b,Yoshii08}
(see Fig.\ \ref{compstrucfig}a) featuring characteristic
triangular bilayers of trigonal-bipyramidal-coordinated Fe as
the electronically active subunit (c.f.\ Fig.\
\ref{compstrucfig}c). Magnetism of the Fe spins attracted an
initial wave of research into these materials, with unusual
features such as anomalous thermomagnetization \cite{Iida86} or
giant coercivity \cite{Iida87} discovered. In all investigated
compounds, a very large Ising anisotropy with spins pointing
perpendicular to the layers was observed
\cite{Iida87,Sugihara78}. The average valence of the Fe ions is
$2.5+$, and therefore Fe$^{2+/3+}$ valence order or CO may also
be expected, first investigated by M{\"o}ssbauer spectroscopy
\cite{Tanaka84}. Hence, there are two binary degrees of freedom
at the Fe sites: Ising spin $\uparrow/\downarrow$ and valence
$2+/3+$.

\paragraph{Geometrical Frustration}

The arrangement of Fe ions leads to ``geometrical frustration''
\cite{Moessner06} hampering both spin order (SO) and CO,
illustrated in Fig.\ \ref{compstrucfig}c) for the Ising spins
by a blue triangle: repulsive or antiferromagnetic (AF)
nearest-neighbor (NN) interactions within the layers can be
satisfied only for two of the spin pairs of the triangle. For
an isolated layer, this would lead to a macroscopic degeneracy
of ground states. The proximity of another triangular layer
changes the topology of the problem, particularly since the
inter-layer Fe-Fe NN-distance (marked $U_2$ in Fig.\
\ref{compstrucfig}c) is about $8\%$ shorter than the
intra-layer one (marked $U_1$). However, there is potential for
frustration in the inter-layer interaction as well, because
each Fe ion has three NN in the other layer (see also Fig.\
\ref{APB}). Furthermore, the rhombohedral stacking leads to
frustration of the interactions between different bilayers,
hindering full 3D order. Even when macroscopic ground-state
degeneracy is broken, geometrical frustration often
\cite{Moessner06} leads to a competition of different phases
with nearly the same energy, facile creation of defects, and
complex unusual ground states, such as the proposed
\cite{Ikeda05} ferroelectric CO in these materials, discussed
in Sec.\ \ref{COsec}.

\paragraph{Strong impact of oxygen stoichiometry}

\begin{figure}[t]%
\sidecaption
\includegraphics*[width=0.55\linewidth]{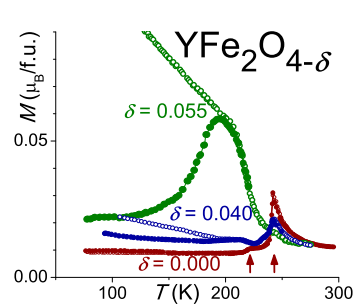}
\caption{%
  Magnetization $M(T)$ of polycrystalline YFe$_2$O$_{4-\delta}$ with different $\delta$, measured in $0.4\, {\rm T}$
  upon heating, after the sample has been cooled in the same field ($\circ$) or in zero-field ($\bullet$). Two transitions in the most stoichiometric sample are marked. Data from \cite{Inazumi81}.}
\label{Inazumi}
\end{figure}

\begin{figure}[t]%
\includegraphics*[width=\linewidth]{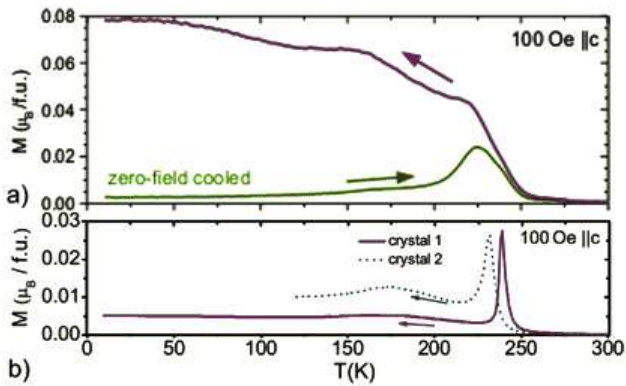}
\caption{%
  Magnetization $M(T)$ of three different LuFe$_2$O$_4$ crystals measured in $100\, {\rm Oe}\|c$, upon cooling except where noted.
  Panel a) after \cite{Phan09}, panel b) after \cite{DeGroot12}.}
\label{LFOmagnChar}
\end{figure}

A soon observed experimental difficulty in these compounds was
a strong variation of physical properties for different
samples. For YFe$_2$O$_{4-\delta}$ this was clearly linked to
tiny variations in oxygen-stoichiometry \cite{Inazumi81}, as
shown in Fig.\ \ref{Inazumi}: magnetization $M(T)$ for the most
off-stoichiometric sample shows a large difference between
zero-field-cooled and field-cooled $M$ suggesting ``glassy''
magnetism without establishment of long-range SO, whereas
$M(T)$ for the stoichiometric sample indicates two distinct
well-ordered antiferromagnetic phases. Single-crystal neutron
diffraction confirmed a tendency towards 3D long-range order
for samples closer to stoichiometry \cite{Funahashi84}, but due
to the absence of sufficiently stoichiometric crystals until
recently \cite{MuellerDipl}, the magnetic structures remain
unsolved. Such a sample-dependence was observed also for Er
\cite{Iida90} and Lu (see Fig.\ \ref{LFOmagnChar}), although
the link to O-stoichiometry \cite{Wang13} is less clear, mostly
due to the range of stoichiometries for which long-range order
is observable shrinking with decreasing $R^{3+}$ size (c.f.\
\cite{Iida90} and Fig.\ \ref{compstrucfig}b). In a wider range
of stoichiometries, links to the remanent magnetization and
resistivity are clear, however \cite{Michiuchi09}.

\paragraph{Ion-size effects and structural modifications}

Changing the $R^{3+}$ size leads to an expected change of the
cell volume ($\sim\! 5\%$ between In and Y), but more relevant
for CO and SO is a $15\%$ change in the ratio between the
layer-separation within a bilayer, $d_b$, and the intra-layer
Fe-Fe distance $a_{\rm hex}$ (Fig.\ \ref{compstrucfig}b), which
can be expected to modify the relative importance of
interactions $U_1$ and $U_2$. Partial substitution, e.g.\ by
Co, Cu, or Mn, on the Fe site generally leads to decreased
conductivity, but also suppressed magnetic ordering (see e.g.\
\cite{Yoshii07,Hou12}). However, effects due to substitution
can be difficult to disentangle from effects due to a
potentially changed O-stoichiometry. Finally, a larger
structural modification can be done by intercalating
$R$Fe$_2$O$_4$ with one or more blocks of $R$FeO$_3$, each
adding an additional $R-$ and a single-Fe layer between the
bilayers \cite{Kato76}. There have been relatively few studies
on these intercalated compounds (e.g.\ \cite{Qin09}). Finally,
a high-pressure orthorhombic modification with a large
supercell, possibly CO, has recently been reported
\cite{Rouquette10}, but the details of this structure have yet
to be elucidated.

\section{Anomalous dielectric dispersion and pyroelectric currents}
\label{macroFE}

\begin{figure}[t]%
\includegraphics*[width=0.85\linewidth]{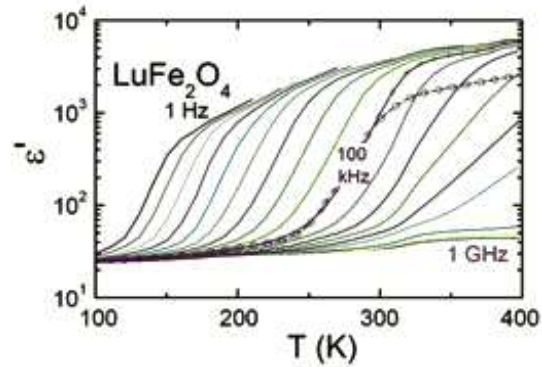}
\caption{%
  Real part of the dielectric permittivity $\varepsilon'$ vs $T$ (lines) in LuFe$_2$O$_4$ for frequencies between $1\, {\rm Hz}$
  and $1\, {\rm GHz}$ equally spaced with two frequencies per decade. The data additionally marked with symbols ($\circ$,$+$) are
  both taken at $100\, {\rm kHz}$, but with electrodes from different material: Graphite ($\circ$) and silver ($+$ and all other
  data). Figure from \cite{Niermann12}, $\copyright \ $ 2012 American Physical Society.}
\label{dielec}
\end{figure}

Many rare earth ferrites, e.g.\ ErFe$_2$O$_4$ \cite{Ikeda94},
have been studied extensively by dielectric spectroscopy, with
anomalously large real parts of the dielectric constant
$\varepsilon'$ in the low-frequency limit generally observed. A
typical (recent) example of such a measurement is shown in
Fig.\ \ref{dielec}. From the frequency-dependence of the
temperature of maximum rise of $\varepsilon'(T)$ a connection
of the dispersion to electrons hopping between Fe$^{2+}$ and
Fe$^{3+}$ has been concluded and an origin of the large
dielectric constants in motion of ferroelectric domain
boundaries suggested \cite{Ikeda05,Ikeda94}. Within this
interpretation, giant ($>\! 20\%$) room-temperature
magneto-dielectric response \cite{Subramanian06} would then
suggest multiferroicity with large magnetoelectric coupling,
potentially useful for applications. Research on possible
ferroelectricity in rare earth ferrites focused on
LuFe$_2$O$_4$, because in this compound superstructure
reflections were found, which for example by resonant x-ray
diffraction at the Fe $K-$edge could be clearly associated with
Fe$^{2+/3+}$ charge order \cite{Ikeda05,Mulders09}, allowing
the discussion of potentially polar CO phases (see Sec.\
\ref{COsec}).

\begin{figure}[t]%
\includegraphics*[width=\linewidth]{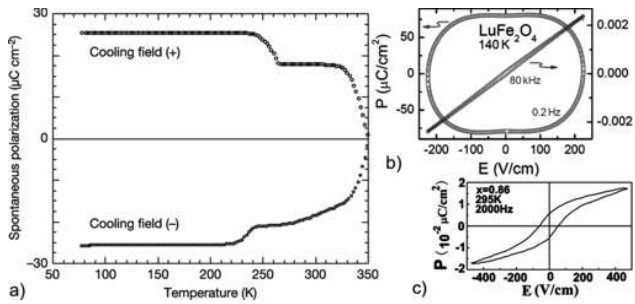}
\caption{%
  a) Electric polarization $P$ vs temperature $T$ of LuFe$_2$O$_4$ as deduced from pyroelectric current measurements
  conducted on warming after the sample had been cooled in electric fields $E\|c$ of $\pm 10\, {\rm kV/cm}$. Panel from
  \cite{Ikeda05}, $\copyright \ $ 2005 Nature Publishing Group. b) $P$ vs $E$ hysteresis loops of LuFe$_2$O$_4$ at $140\, {\rm K}$
  for two different frequencies, from \cite{Ruff12}, $\copyright \ $ EDP Sciences, Societ\`{a} Italiana di Fisica, Springer-Verlag
  2012. With kind permission of The European Physical Journal (EPJ). c) $P(E)$ of Lu$_2$Fe$_{2.14}$Mn$_{0.86}$O$_7$ at $2\,{\rm kHz}$,
  from \cite{Qin09}, $\copyright \ $ 2009 American Institute of Physics.}
\label{PE}
\end{figure}

Macroscopic proof of ferroelectricity requires demonstration of
a remanent electric polarization, which should be switchable by
application of an electric field. A remanent polarization
larger than in the traditional ferroelectric BaTiO$_3$ was
suggested by pyroelectric current measurements \cite{Ikeda05},
shown in Fig.\ \ref{PE}a). These measurements were performed by
first cooling the single-crystal sample in an electric field of
$\pm 10\,{\rm kV/cm}$, then switching off the electric field,
and finally measuring, upon warming in $E=0$, the pyroelectric
current associated with the change of the polarization. The
dependence of the pyroelectric current and thus the
polarization on the direction of the cooling field is
consistent with a macroscopic polarization, and the decay to
zero around the charge ordering temperature of $T_{CO}\sim
320-330\,{\rm K}$ \cite{Yamada97,Angst08} supports a connection
of $P$ with the charge ordering, although the lack of a
saturation region with no pyrocurrent flowing anymore above
$T_{CO}$ renders this last conclusion tentative. It was noted,
however, that similar behavior can also be observed in
(non-ferroelectric) leaky dielectrics, where it originates from
localization of free carriers at interfaces \cite{Maglione08}.

Subsequent measurements by various groups partly reproduced the
pyrocurrent results of Fig.\ \ref{PE}a), as noted e.g.\ in
\cite{Mulders09}, or tried to establish $P(E)$ hysteresis loops
at low temperatures to more directly show ferroelectricity. Out
of many attempts at the latter, only few were published
\cite{Ruff12,Park07,Viana11}, some showing hysteresis, but not
the saturation necessary \cite{Scott08} to establish
polarization switching. Fig.\ \ref{PE}b) shows $P(E)$ measured
at $140\,{\rm K}$ and two frequencies \cite{Ruff12}: the
low-frequency loop shows substantial hysteresis, but no
indications of saturation, whereas the loop measured with a
frequency of $80\, {\rm kHz}$ exhibits practically linear
behavior without appreciable hysteresis.

Because the polarization is linked with the dielectric
constant, $\mathbf{P}=(\varepsilon -1)\varepsilon_0
\mathbf{E}$, detailed modeling of the dielectric dispersion can
show the origin of observed polarization behaviors. This was
done for the results obtained on LuFe$_2$O$_4$ single crystals
over 9 decades of frequencies \cite{Niermann12} and shown in
Fig.\ \ref{dielec}. The qualitative behavior with large
$\varepsilon'$ at low frequency and high $T$ is the same as in
previous observations \cite{Ikeda05}, but a strong impact of
the electrode material as shown for the example of $100\,{\rm
kHz}$ (different symbols) already suggests significant contact
effects, e.g.\ due to Schottky-type depletion layers at the
contact interfaces, which act as an additional capacitance
\cite{Niermann12}. The whole data-set of $\varepsilon'(f,T)$
and $\varepsilon''(f,T)$ could be well described with an
equivalent-circuit model featuring contact-capacitance and
-conductance, and sample hopping- and dc-conductivity, in
addition to the intrinsic dielectric constant $\varepsilon_i$
of the material, all parameters independent of frequency. The
fitted $\varepsilon_i (T)$ is only around $30$ and does not
show any indications of ferroelectric or antiferroelectric
transitions. Two recent studies on polycrystalline
LuFe$_2$O$_4$ reached the same conclusion \cite{Ruff12,Ren11}.
For low frequencies or high $T$, the circuit reduces to a leaky
contact capacitance in series with the sample resistivity,
resulting in large $\varepsilon'$. Because the step in
$\varepsilon'$ mainly depends on contact capacitance and sample
resistance \cite{Niermann12} a magneto-resistance will lead to
an apparent magnetodielectric response, according to the
analysis in \cite{Ren11} the latter can be fully accounted for
by the former. A very recent study \cite{Kambe13} of dielectric
spectroscopy using various contact materials suggests a small
($\sim\! 1\%$) sample magnetodielectric effect at least at
$220\, {\rm K}$ (at the AFM-fM transition, c.f.\ Fig.\
\ref{NDH}), but the simultaneously observed 18 times larger
change in resistivity may influence the fitted sample capacity.

$P(E)$ loops can only show the intrinsic behavior if measured
at frequencies and temperatures where $\varepsilon'(f,T)$ is
close to $\varepsilon_i (T)$. For the measurements shown in
Fig.\ \ref{PE}b), this is the case for $P(E)$ measured at
$80\,{\rm kHz}$, but clearly not for $P(E)$ measured at
$0.2\,{\rm Hz}$. There is one $P(E)$ loop in the literature
\cite{Qin09} that {\em does} seem to indicate ferroelectricity
given the expected tendency to saturate in high $E$ (Fig.\
\ref{PE}c), measured on intercalated Lu$_2$Fe$_3$O$_7$, in
which conductivity has been suppressed by $\sim\! 29\%$ Mn for
Fe substitution. The occurrence of at least piezoelectricity in
this compound is supported by piezoresponse force microscopy
\cite{Chen12}. Although the switchable polarization is very
weak (compare panels c and a), smaller even than in
``spin-spiral multiferroics'' like TbMnO$_3$, this suggests
that intercalated compounds deserve more attention than they
currently get.

\begin{figure}[bt]%
\includegraphics*[width=\linewidth]{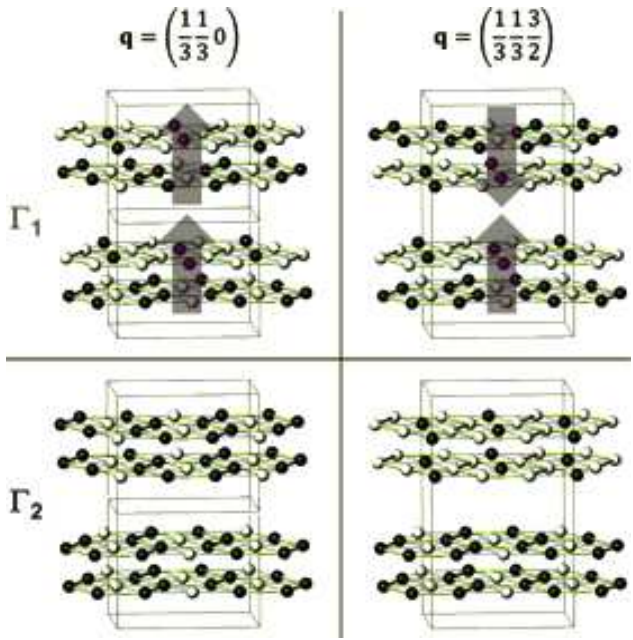}
\caption{%
  Sketch of the $4$ symmetry-allowed Fe$^{2+}$ (black) / Fe$^{3+}$ (white)
  charge configurations following ($\frac{1}{3}\frac{1}{3}0$) or ($\frac{1}{3}\frac{1}{3}\frac{3}{2}$) propagation
  (with additional contribution by ($000$) or ($00\frac{3}{2}$), respectively). Arrows indicate bilayer polarization.}
\label{COs}
\end{figure}

\section{Charge Order}
\label{COsec}

It was the combination of macroscopic indications of
ferroelectricity, discussed in Sec.\ \ref{macroFE}, with a
likely microscopic model of charge order (CO) that is
electrically polar, which made a convincing case for
LuFe$_2$O$_4$ exhibiting ferroelectricity from CO
\cite{Ikeda05}. This CO-model was deduced by scattering
methods, in which signatures of the order in reciprocal space
are investigated.

\begin{figure*}[t]%
\includegraphics*[width=\linewidth]{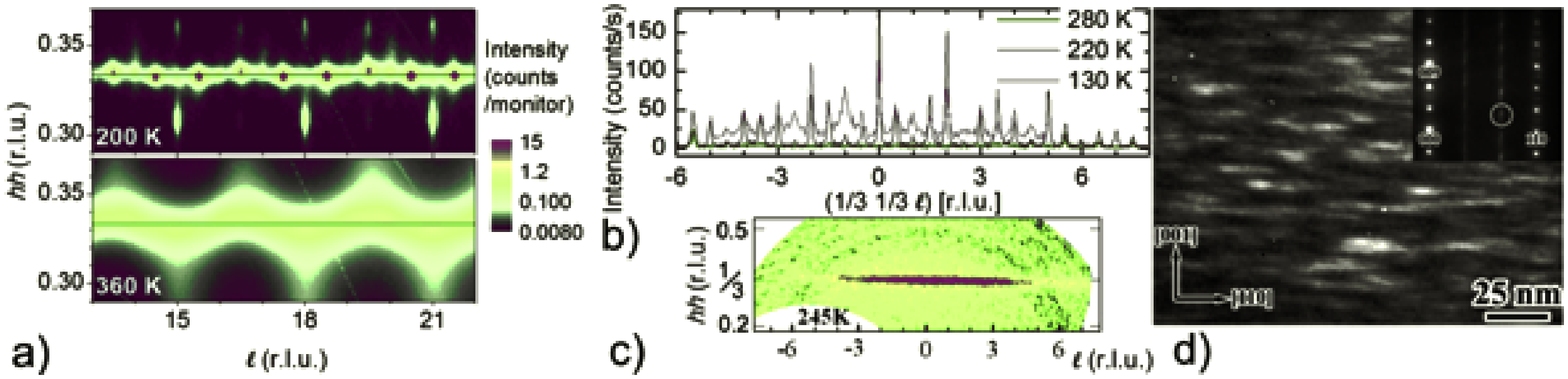}
\caption{%
  Charge order CO superstructure in LuFe$_2$O$_4$ by different techniques. a) X-ray diffraction scattered intensity at ($hh\ell$) at
  $200$ and $360\, {\rm K}$. From \cite{Angst08}, $\copyright \ $ 2008 American Physical Society. b) Neutron diffraction scans
  along ($\frac{1}{3}\frac{1}{3}\ell$) at three temperatures. From \cite{Christianson08}, $\copyright \ $ 2008 American
  Physical Society. c) Neutron diffraction intensity in spin-flip channel at ($hh\ell$) at $245\, {\rm K}$. From \cite{DeGroot12}, $\copyright \ $ 2012 American Physical Society. d) Electron-diffraction dark-field image taken at $92\, {\rm K}$ with a ($\frac{1}{3}\frac{1}{3}\frac{5}{2}$)
  superstructure spot, circled in the inset. The CO domains extend $\sim\! 30\, {\rm nm}$ in the $ab$-plane and $\sim\! 6\, {\rm nm}$
  along $c$. Inset: electron diffraction pattern with ($1\overline{1}0$) incidence. From \cite{Park09}, $\copyright \ $ 2009
  American Physical Society.}
\label{XRDND}
\end{figure*}

\begin{figure}[b]%
\includegraphics*[width=0.96\linewidth]{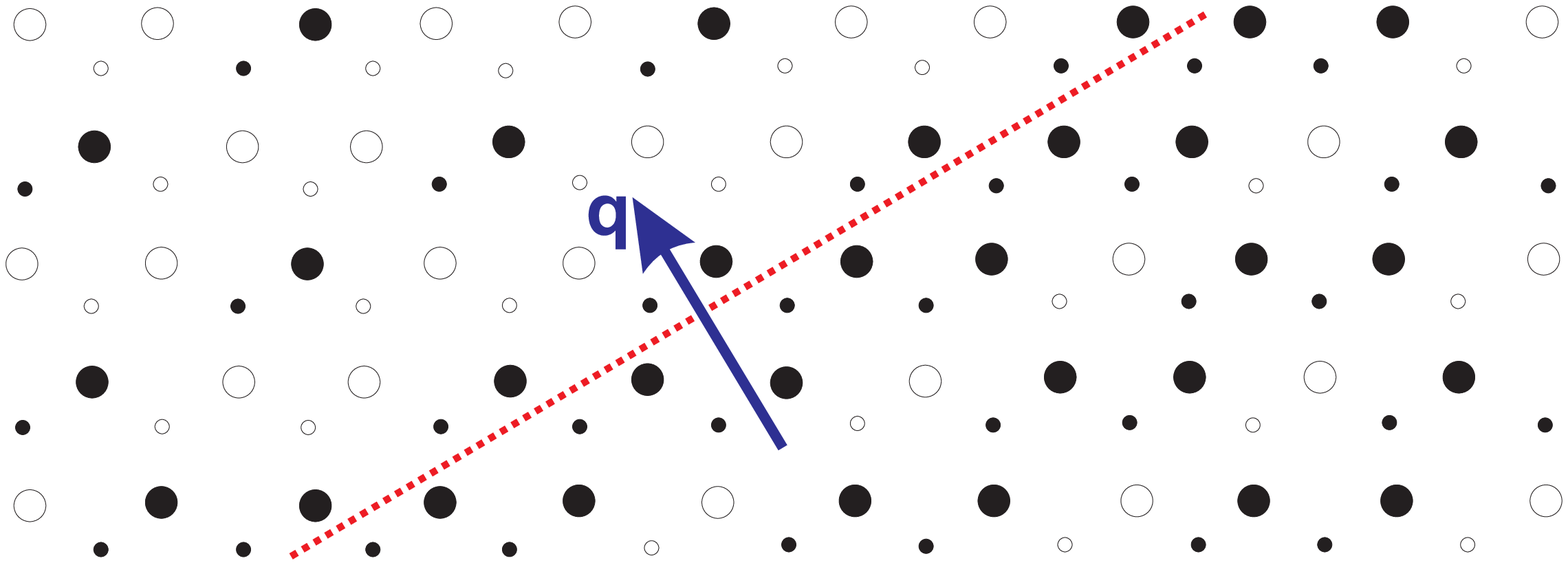}
\caption{%
  Example of a possible anti-phase-boundary (dashed) perpendicular
  to the in-plane propagation of CO, which switches the majority valence. Top-view of bilayer ($\bullet$=Fe$^{2+}$,$\circ$=Fe$^{3+}$), Fe ions in the lower layer are drawn smaller.}
\label{APB}
\end{figure}

For many $R$Fe$_2$O$_4$, at least diffuse scattering of x-rays
at the ($\frac{1}{3}\frac{1}{3}\ell$) line in reciprocal space
(in hexagonal-cell notation) suggested a tendency towards a CO
that could be described, for example, with a so-called
$\sqrt{3}\times\sqrt{3}$ cell. Alternatively, a propagation
vector with ($\frac{1}{3}\frac{1}{3}$) in-plane component can
be conveniently described by expressing the crystal structure
without CO in a $C-$centered monoclinic cell, which is then
enlarged three times along its $b-$axis. Because such a CO
breaks the 3-fold rotation symmetry of the crystal structure,
domains of CO with symmetry-equivalent
($\frac{\overline{2}}{3}\frac{1}{3}$) and
($\frac{1}{3}\frac{\overline{2}}{3}$) propagations and
$120^{\circ}$ rotated monoclinic cells, as sketched in Fig.\
\ref{compstrucfig}, also exist \cite{Angst08}.

The simplest arrangement of Fe$^{2+}$ and Fe$^{3+}$ ions in a
bilayer leading to such propagations and maintaining charge
neutrality is the one proposed in \cite{Ikeda05} and shown in
Fig.\ \ref{COs} top-left (two cells of the CO are shown). Here,
for each bilayer, the upper layer has a surplus of Fe$^{3+}$
ions and the lower layer a surplus of Fe$^{2+}$ ions, which
implies an electric dipole moment between the layers (indicated
by arrows), i.e.\ this CO is indeed ferroelectric, rendering
the bilayers polar.

\subsection{Long-range CO from diffraction}
\label{LRCO}

In the case of LuFe$_2$O$_4$, actual superstructure reflections
were observed, below $T_{\rm CO}\sim 320\,{\rm K}$, by several
groups using various diffraction techniques (Fig.\
\ref{XRDND}). Most studies resolving 3D peaks found the
principal superstructure reflections to be near positions
($\frac{1}{3},\frac{1}{3},$halfinteger) and symmetry-equivalent
\cite{Ikeda05,DeGroot12b,Yamada00,Mulders09,Yamada97,Angst08,Christianson08,Park09,Ikeda08,Wen09,Wen10,Mulders11,Bartkowiak12,Bourgeois12},
with in-plane/out-of-plane correlation lengths given as e.g.\
$30$/$6\,{\rm nm}$ \cite{Park09} or $80$/$7\,{\rm nm}$
\cite{Wen09}. Resonant x-ray diffraction at the Fe K-edge of
these reflections support their association with Fe$^{2+/3+}$
CO \cite{Ikeda05,Mulders09,Ikeda08,Bartkowiak12}. Sharp
reflections at other positions, including near
($\frac{1}{3},\frac{1}{3},$integer)
\cite{Zhang07,Zhang07b,Yang10b} are sometimes observed in
electron diffraction studies, generally on samples with higher
oxygen content \cite{Bourgeois12,Yang10b}. The additional
reflections in \cite{Bourgeois12} were attributed to ordering
of excess oxygen ions.

\begin{figure*}[t]%
\sidecaption
\includegraphics*[width=.62\linewidth]{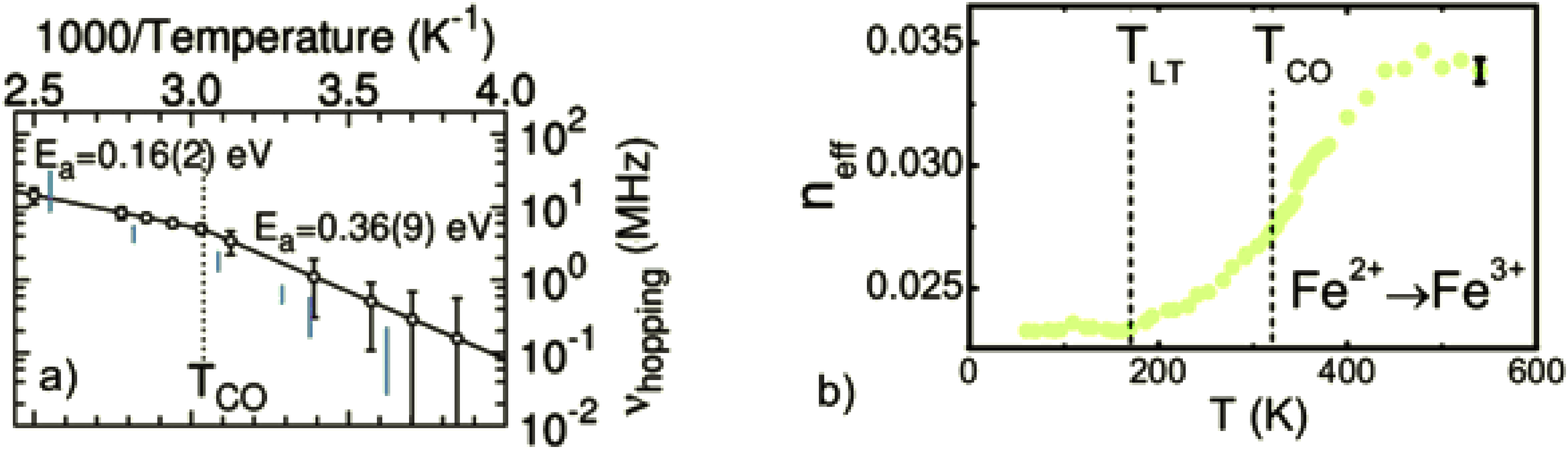}
\caption{%
  Charge dynamics in LuFe$_2$O$_4$. From \cite{Xu08}, $\copyright \ $ 2008 American Physical Society. a) Arrhenius-plot of the ${\rm Fe}^{2+}\!\rightarrow{\rm
Fe}^{3+}$ hopping frequency determined from M{\"o}ssbauer
spectra ($\circ$, bars are from \cite{Tanaka84}). b)
Temperature-dependence of the strength $n_{\rm eff}$ of an
optical excitation assigned to ${\rm Fe}^{2+}\rightarrow{\rm
Fe}^{3+}$ charge transfer.} \label{spectroscopy}
\end{figure*}

High-resolution x-ray diffraction
\cite{Yamada00,Yamada97,Angst08} found the exact positions of
the ($\frac{1}{3},\frac{1}{3},$halfinteger) reflections to be
well described by $\mathbf{s}+\mathbf{p_A}$,
$\mathbf{s}+\mathbf{p_B}$, and $\mathbf{s}+\mathbf{p_C}$, where
$\mathbf{s}$ is a structural reflection following the
rhombohedral centering condition $-h+k+\ell=3n$ with $n$
integer. The three symmetry-equivalent propagation vectors,
corresponding to CO domains, are
$\mathbf{p_A}=(\frac{1}{3}+\delta,\frac{1}{3}+\delta,\frac{3}{2})$,
$\mathbf{p_B}=(\frac{\overline{2}}{3}-2\delta,\frac{1}{3}+\delta,\frac{3}{2})$,
and
$\mathbf{p_C}=(\frac{1}{3}+\delta,\frac{\overline{2}}{3}-2\delta,\frac{3}{2})$
and $\delta\sim 0.003$ $T-$dependent \cite{Angst08}. A
secondary set of reflections is described by
$\mathbf{p'_A}=(\tau,\tau,\frac{3}{2})$,
$\mathbf{p'_B}=(\overline{2\tau},\tau,\frac{3}{2})$, and
$\mathbf{p'_C}=(\tau,\overline{2\tau},\frac{3}{2})$, and
$\tau\sim 9\delta$ following the $T-$dependence of $\delta$.
Although this seems to suggest an incommensurate modulation of
the Fe valence state, there is no evidence of a ``true''
incommensuration, which would imply a wide distribution of Fe
valence states from $2+$ to $3+$, in conflict with
spectroscopic measurements \cite{Tanaka84,Xu08,Ko09}. Much more
likely is therefore a locally commensurate state, interspersed
with discommensurations or anti-phase-boundaries (Fig.\
\ref{APB}) of an average separation, estimated from
$\tau\!\sim\! 0.028$ \cite{Angst08}, of about $12\, {\rm nm}$.
It has been demonstrated on other charge-ordering oxides that a
proliferation of such anti-phase-boundaries, which can cost
very little energy due to geometrical frustration, can lead to
sharp superstructure reflections at incommensurate positions
\cite{Fe2OBO3second}.

Consequently, the possible CO corresponding to
($\frac{1}{3}\frac{1}{3}\frac{3}{2}$) and for completeness also
($\frac{1}{3}\frac{1}{3}0$) have to be examined through
symmetry-analysis \cite{Bertaut68}. For each of these
propagation vectors, there are two irreducible representations,
corresponding to the two Fe ions of the primitive unit cell
(labeled $\texttt{1}$ and $\texttt{2}$ in Fig.\
\ref{compstrucfig}a) having the same ($\Gamma_2$) or different
($\Gamma_1$) valence. Each corresponding CO contains more than
than two valence states, in contradiction to the results of
M{\"o}ssbauer spectroscopy, which imply a bimodal valence
distribution \cite{Xu08}. Such a bimodal distribution is
obtained by combining, in a unique way \cite{Angst08}, the CO
from ($\frac{1}{3}\frac{1}{3}\frac{3}{2}$) [resp.\
($\frac{1}{3}\frac{1}{3}0$)] with the one from
($00\frac{3}{2}$) [resp.\  ($000$)].

The four resulting CO are shown in Fig.\ \ref{COs}. The CO with
$\Gamma_2$ (lower row) have bilayers with a net charge, indeed
for ($\frac{1}{3}\frac{1}{3}0$)/($000$) the whole structure
would be charged. ($\frac{1}{3}\frac{1}{3}0$)/($000$) with
$\Gamma_1$ gives the ferroelectric CO with polar bilayers
proposed in \cite{Ikeda05}. However, since the observed (also
in \cite{Ikeda05}) superstructure reflections follow
($\frac{1}{3}\frac{1}{3}\frac{3}{2}$) propagation, one of the
two CO shown on the right of Fig.\ \ref{COs} should be
realized, either (top) the antiferroelectric CO with polar
bilayers but zero net polarization, as proposed in
\cite{Angst08}, or (bottom) the CO with stacking of oppositely
charged bilayers. The latter was initially dismissed because it
involves a transfer of charge between neighboring bilayers
separated by $\sim 6\,{\rm {\AA}}$, but an experimental
determination of which CO is realized requires a full
structural refinement.

This was recently achieved \cite{DeGroot12b}, the key being the
screening of a large quantity of small crystals of the highest
quality, as judged by $M(T)$ (Fig.\ \ref{LFOmagnChar}), for one
with almost only one of the three possible CO-domains
populated. The structure obtained at $210\,{\rm K}$, readily
refined in the space group $C2/m$ with a residual of $5.96\%$,
contains four iron-sites with valences, determined with the
bond-valence-sum (BVS) method \cite{Brown02}, $1.9$, $2.1$,
$2.8$, and $2.9$ \cite{DeGroot12b}. Given the uncertainties
inherent to the empirical BVS-method, these values can be
considered to be remarkably close to the ideal values of $2$
and $3$ (for a comparison of other charge ordering ferrites see
Fig.\ 5 in \cite{Angst07}). The thus identified CO is the one
with charged bilayers, shown in Fig.\ \ref{COs} bottom right.
Attempts to force-refine the other CO or to refine in a
lower-symmetry space group confirmed this structure solution,
and independent confirmation was also obtained indirectly via
spin-charge coupling (c.f.\ Sec.\ \ref{SCC}). The ground-state
CO of LuFe$_2$O$_4$ is thus not ferroelectric and does not
contain the initially proposed polar bilayers.

\subsection{Short-range correlations and charge dynamics}
\label{SRCO}

\begin{figure*}[bt]%
\includegraphics*[width=\linewidth]{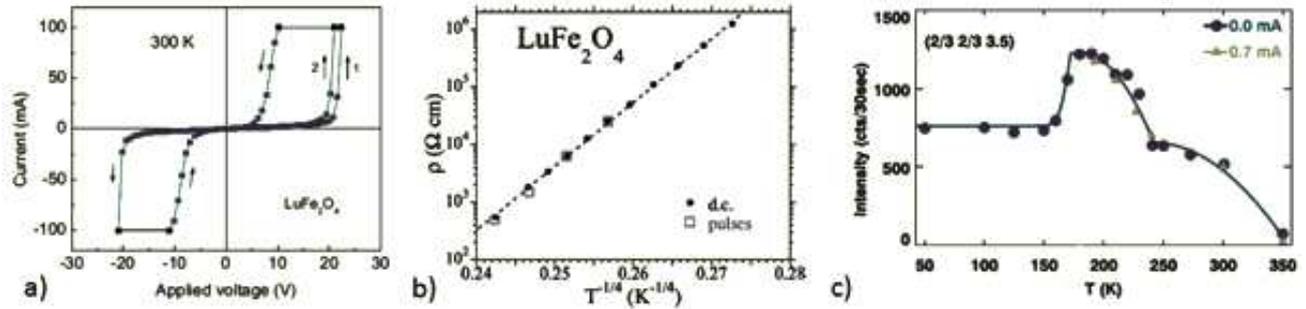}
\caption{%
  Electric-field effects in LuFe$_2$O$_4$. a) Current-voltage hysteresis loop at $300\, {\rm K}$, with the current limited
  to $100\, {\rm mA}$. From \cite{Li08}, $\copyright \ $ 2008 American Institute of Physics. b) Resistivity $\rho$ vs $T^{-1/4}$.
  full symbols: low $E$-field d.c. measurements; open symbols: high-field pulsed measurements. From \cite{Fisher11},
  $\copyright \ $ 2011 American Institute of Physics. c) Neutron diffraction intensity of
  ($\frac{2}{3}\frac{2}{3}\frac{7}{2}$) as a function of calibrated temperature, obtained by cooling with $0.7\, {\rm mA}$
  ($\triangle$) or no current ($\bullet$). From \cite{Wen10}, $\copyright \ $ 2010 American Physical Society.}
\label{IV}
\end{figure*}

Above $T_{\rm CO}$, the sharp CO superstructure reflections in
LuFe$_2$O$_4$ are replaced by diffuse scattering along
$(\frac{1}{3}\frac{1}{3}\ell)$, with a characteristic zig-zag
pattern, shown in Fig.\ \ref{XRDND}a) lower graph, discernable
up to about $550\,{\rm K}$ \cite{Yamada97,Zhang07b}. A detailed
analysis of the scattered intensity \cite{Angst08} suggests
that it can be described as very broad and overlapping peaks at
positions of structural reflections $\pm
(\frac{1}{3}-\delta',\frac{1}{3}-\delta',0)$ and
symmetry-equivalent. Corresponding to the large peak-width
along $\ell$, at $360\,{\rm K}$ a correlation extending only to
$2-3$ bilayers was concluded, whereas the remaining relative
sharpness along ($hh0$) suggests that in each individual
bilayer, medium-range CO is maintained much above $T_{\rm CO}$.
Applying symmetry-analysis on the positions of the diffuse
peaks suggests a marginal tendency towards ferro-stacking of
neighboring bilayers, one of the CO shown in Fig.\ \ref{COs}
left. In \cite{Angst08}, representation $\Gamma_1$ and thus a
tendency towards ferroelectric CO, was assumed. However, as it
is now clarified that in the long-range ordered state,
representation $\Gamma_2$ with charged rather than polar
bilayers is realized, it seems more likely that the short-range
CO is also following $\Gamma_1$, without polar bilayers ever
being present.

This seems at first impossible, as this corresponds to an
overall net charge of the structure. However, given the small
correlation-volume deduced from the broadness of the diffuse
peaks, a charge-transfer over a few bilayers (via a presently
unknown pathway) is sufficient, and feasible considering that
this is not a static, but a dynamic short-range CO, i.e.\
short-range-correlated valence fluctuations. M{\"o}ssbauer
spectroscopy can directly assess the dynamics of CO and charge
fluctuations: spectra of several $R$Fe$_2$O$_4$ were analyzed
with a model featuring isomer shifts, quadrupole splittings,
and linewidths for both valence states and additionally a ${\rm
Fe}^{2+}\!\rightarrow{\rm Fe}^{3+}$ hopping frequency, showing
still discernable natural hopping even somewhat below $T_{\rm
CO}$ \cite{Tanaka84}.

Fig.\ \ref{spectroscopy}a) shows the hopping frequency
$\nu_{\rm hopping}$ determined in a study on powdered crystals
of optimal quality. The hopping above $T_{\rm CO}$ can be
described by an Arrhenius law with an activation energy of
$\sim\! 0.16\,{\rm eV}$. $\nu_{\rm hopping}$ decreases
continuously upon cooling through $T_{\rm CO}$, and below can
again be described by an Arrhenius law, but with a higher
activation energy $\sim\! 0.36\,{\rm eV}$. The Arrhenius law
with a change of slope at $T_{\rm CO}$ is consistent with the
$T-$dependence of the dc conductivity extracted by dielectric
spectroscopy on similar samples \cite{Niermann12}. The
substantial electron hopping present even below $T_{\rm CO}$,
where diffraction indicates long-range CO, likely involves
fluctuations of anti-phase boundaries, a scenario shown to be
at work in another CO ferrite, and connected to $T-$dependent
apparent incommensuration \cite{Fe2OBO3second}. In
LuFe$_2$O$_4$ the incommensuration becomes constant only below
$\sim\! 170\,{\rm K}$, suggesting that the charge dynamics
become completely frozen only at temperatures substantially
below $T_{\rm CO}$.

The capacity of ${\rm Fe}^{2+}\!\!\rightarrow\! {\rm Fe}^{3+}$
charge transfer can be probed not only through natural electron
hopping, but also by driving the process at optical
frequencies. Fig.\ \ref{spectroscopy}b) shows the
$T-$dependence of the strength of an optical excitation
assigned to ${\rm Fe}^{2+}\!\!\rightarrow\! {\rm Fe}^{3+}$
charge transfer \cite{Xu08}. Upon cooling, it starts to
decrease around $500\,{\rm K}$, reasonably close to the
reported \cite{Yamada97,Zhang07b} onset of diffuse scattering,
and it bottoms out only at $T_{\rm LT}\sim\! 170\,{\rm K}$,
consistent with the freezing of charge dynamics at this
temperature. The nature of the phase below $T_{\rm LT}$ will be
further discussed in Sec.\ \ref{LT}.

\subsection{Electric-field and current effects on CO}
\label{COinE}

Given that the pyroelectric current measurements shown in Fig.\
\ref{PE}a) had been prepared by first cooling the sample in an
electric field, and that diffraction analyses failed to deduce
a ferroelectric CO, it had been proposed \cite{Angst08} that
such a ferroelectric CO might be stabilized by electric fields
$E$ or by cooling in $E$. However, at low $T$ electric fields
up to $20\,{\rm kV/cm}$ were shown to have no effect on the CO
\cite{Wen10}. At higher $T$, intriguing electric-field effects
have been observed macroscopically in the form of non-linear
current-voltage characteristics (Fig.\ \ref{IV}a)
\cite{Li08,Ikeda11}: Rather than eventually reaching a state
with lower resistance as might be expected for a ferroelectric
phase, a current-breakthrough is observed. The breakthrough
depends on the environment so that a possible application in
gas-sensing has been proposed \cite{Cao12}. Effects were also
reported on dielectric constants \cite{Li08b} and magnetism
\cite{Mulders11,Li09,Wang10}. With in-situ electron diffraction
experiments, a disappearance of the CO superstructure
reflections at this threshold was observed \cite{Zeng08,Cao11},
suggesting an electric-field or current-induced melting of the
CO. In contrast, an ex-situ in-field cooling experiment
suggests improved CO correlations, without affecting the
intensity-distribution between different reflections
\cite{LawrencePhD}.

However, current-voltage measurements using single short
($\sim\! 1\,{\rm ms}$) current pulses \cite{Fisher11} found
Ohmic behavior in polycrystalline LuFe$_2$O$_4$ (see Fig.\
\ref{IV}b) up to $E$ much higher than fields where the
breakdown in dc measurements occurs, with the time-dependence
on longer pulses suggesting self-heating effects
\cite{Fisher11}. Similar pulsed measurements on YbFe$_2$O$_4$
\cite{Wang12} suggest non-linearity in the in-plane
conductivity, attributed to sliding charge density and possibly
a break-down of CO, whereas large self-heating effects for
longer pulses are also found. An in-situ neutron diffraction
experiment \cite{Wen10} demonstrates the absence of any
current-dependence on either charge or magnetic order when the
sample-temperature is properly recalibrated (Fig.\ \ref{IV}c).
A CO unaffected by static (melting of the CO by $1.55\,{\rm
eV}$ photons was demonstrated by a recent pump-probe experiment
\cite{Itoh13}) electric fields or currents was also found by
in-situ infrared spectroscopy \cite{Vitucci11}, by in-situ
x-ray diffraction with the real sample temperature established
from the thermal expansion of the silver electrodes
\cite{DeGrootPhD}, and by resonant x-ray diffraction after
in-field cooling \cite{Bartkowiak12}. Although a recent
infrared-spectroscopy study \cite{Lee12} suggests very subtle
field-effects not attributable to self-heating, the conclusion
that a different, possibly ferroelectric, CO cannot be
stabilized by electric fields is unavoidable.

\subsection{Theoretical considerations}
\label{COtheory}

The experimentally deduced CO with charged bilayers is
surprising, because CO is usually assumed to be driven by the
repulsion of electrons between different sites (``Wigner
crystallization''), which should result in minimizing the
occurrence of neighbors of the same valence. It is clear that
packing electrons close together in a charged bilayer should
lead to a significant energy penalty. Although the full
description of the 3D CO requires at least the four
interactions $U_1-U_4$ indicated in Fig.\ \ref{compstrucfig}c)
\cite{Yamada00,Harris10}, the clear indications by the diffuse
scattering above $T_{\rm CO}$ (Sec.\ \ref{SRCO}) that
correlations within the bilayers are much stronger than those
between different bilayers make a first analysis based on only
$U_1$ and $U_2$ (intrabilayer interactions) useful. Concerning
the intralayer interaction $U_1$ all the CO of Fig.\ \ref{COs},
as well as many others, such as a stripe-like CO, are exactly
degenerate due to geometrical frustration (c.f.\ Sec.\
\ref{family}). However, concerning the three nearest neighbors
(NN), which are in the other layer ($U_2$), the charged bilayer
CO (Fig.\ \ref{COs} bottom) have on average $\frac{5}{3}$
same-valence NN, whereas the polar CO (Fig.\ \ref{COs} top)
have only $\frac{4}{3}$ same-valence NN, and is therefore more
stable for repulsive $U_2$. However, a stripe-like CO can have
$1$ same-valence NN, which should make it even more favorable,
as found also by Monte-Carlo simulations taking into account
the long-range nature of the Coulomb interaction
\cite{Xiang07}. This is an indication that Coulomb repulsion is
not the sole relevant driving force for CO in $R$Fe$_2$O$_4$.
It is not really surprising: a proof of purely
electrostatically driven CO has been elusive in bulk materials
and strong influence of lattice distortions on CO are well
known, e.g.\ in manganites \cite{Popovic02}. In addition to
lattice effects, indicated to be relevant in LuFe$_2$O$_4$ by
infrared \cite{Xu10,Vitucci10} and Raman \cite{Hou11}
spectroscopy, CO driven by magnetic exchange has also been
demonstrated \cite{McQueeney07}.

Density-functional-theory (DFT) calculations performed for
LuFe$_2$O$_4$ \cite{Xiang07} suggest a lattice-contribution of
about $40\%$ to the total energy gain by CO for the
ferroelectric ($402\,{\rm meV/f.u.}$) and stripe ($384\,{\rm
meV/f.u.}$) CO investigated, estimated by separately
stabilizing the CO with fixed atomic positions and with
relaxing the structure. Charged bilayer or antiferroelectric CO
could not be obtained, because a cell with $3$ bilayers was
used. A ferrielectric CO with a $\uparrow\downarrow\uparrow$
stacking of bilayer polarizations (and a net polarization close
to the one indicated in Fig.\ \ref{PE}a, lending support to
this model) and later using a larger cell an antiferroelectric
CO \cite{Angst08} was found to be yet more stable, by a total
of $15\,{\rm meV/f.u.}$. These small energy differences between
antiferroelectric, ferroelectric, and stripe CO are an
indication of the relevance of geometrical frustration, as
expected (Sec.\ \ref{family}). Spin-orbit coupling and
magnetism was considered in a later study \cite{Xiang09} (see
Sec.\ \ref{SCC}). In this study, a fixed ferroelectric CO was
used, but the reported energy gain of $78\,{\rm meV/f.u.}$ is
considerably larger than the above energy-differences between
different CO, indicating that magnetism may be relevant in
stabilizing the CO. The charged bilayer CO has yet to be tested
by DFT.

The CO in $R$Fe$_2$O$_4$ has also been studied by lattice gas
models. Modeling a single bilayer
\cite{Nagano07,Naka08,Nasu08,Watanabe09,Watanabe10} already
yields several competing CO phases, among them the
ferroelectric CO (Fig.\ \ref{COs} top left) and non-polar CO
phases with in-plane propagations such as ($\frac{1}{2}0$),
($\frac{1}{4}\frac{1}{4}$), ($\frac{1}{6}\frac{1}{6}$) and
($\frac{5}{12}\frac{5}{12}$), again pointing out the importance
of geometrical frustration. Because in
\cite{Nagano07,Naka08,Nasu08,Watanabe09,Watanabe10} charge
neutrality of this bilayer was implied, the charged bilayer CO
was not considered. A lattice gas model defined on the whole 3D
lattice of Fe has been considered in \cite{Yamada00,Harris10},
though without considering in detail the charged bilayer CO
found experimentally. Starting point is the Hamiltonian
\begin{equation}
\label{eq1} \mathcal{H}=\sum {U_{ij}^{\nu\nu'}\cdot \sigma_i^\nu \sigma_j^{\nu'}},
\end{equation}
where $i/j$ enumerate the primitive unit cells, $\nu/\nu'=1,2$
denote the Fe ion in the primitive cell, $\sigma$ assumes the
value $1$ ($-1$) for the valence state $2+$ ($3+$) of the Fe
ion, and $U$ denotes screened Coulomb-interactions between
different Fe ions. In reciprocal space, the eigenvalues of the
Fourier-transformed Hamiltonian (\ref{eq1}) give the energies
for charge patterns defined by a corresponding eigenvector and
a propagation vector, and the charge pattern with the lowest
energy corresponds to the realized CO. Four interactions ($U_1
- U_4$ in Fig.\ \ref{compstrucfig}c) are considered. It is
clear that at least interactions $U_1 - U_3$ are needed to
yield a 3D CO: They lead to instabilities at
($\frac{1}{3}\frac{1}{3}\ell$), but these are degenerate along
$\ell$, yielding only 2D CO \cite{Yamada00}. This is due to the
rhombohedral stacking of the bilayers \cite{Harris10}, a
geometrical frustration effect.

With a non-zero $U_4$, depending on the signs of $U_4$ and
$U_2\cdot U_3$ possible ordering wavevectors
($\frac{1}{3}\pm\delta,\frac{1}{3}\pm\delta,\frac{3}{2}$) or
($\frac{1}{3}\pm\delta,\frac{1}{3}\pm\delta,0$) are obtained.
$U_4<0$ (attractive, i.e.\ favoring same valence) must be
assumed to yield the observed propagation vector, which was
proposed to be due to overscreening \cite{Yamada00}. The
observed (c.f.\ Sec.\ \ref{SRCO}) discrepancy between the
dominant fluctuations above $T_{\rm CO}$ with peaks near
($\frac{1}{3},\frac{1}{3},$integer) and the long-range CO below
$T_{\rm CO}$ with reflections near
($\frac{1}{3},\frac{1}{3},$half-integer) is difficult to
reconcile with a mean-field lattice gas model. In
\cite{Harris10}, the coupling to a non-critical mode,
specifically a particular Raman-active phonon-mode, was
proposed. Raman scattering has so far been conducted only on
polycrystalline samples, not finding any indications for this
scenario \cite{Hou11}. However, a problem of the model is that
with the simple quadratic Hamiltonian (\ref{eq1}), an
instability at only a single $\mathbf{q}$-position can be
obtained, i.e.\ the additionally observed reflections near
($00\frac{3}{2}$) cannot be explained.

\subsection{Ion-size effects on CO}
\label{COYFO}

Given the large distortions of the bilayer in $R$Fe$_2$O$_4$ as
a function of $R$ ion-size (Sec.\ \ref{family} and Fig.\
\ref{compstrucfig}b), different CO for different $R$ may be
expected. Yb is quite close in ion size to Lu, and whereas
long-range CO was not yet found, presumably due to
stoichiometry-problems, diffuse scattering near
($\frac{1}{3}\frac{1}{3}\ell$) is generally observed
\cite{Hearmon12}, and an electron-diffraction study
\cite{Murakami07} even found maxima at $\ell$ half-integer.
Given similar spin-charge coupling (see Sec.\ \ref{SCC}), the
same CO as for Lu is likely. For $R$=In, which has a
significantly smaller ion size, single crystals are not
available, but synchrotron x-ray powder diffraction suggests
the appearance of long-range CO below $\sim\! 250\,{\rm K}$,
with peaks indexed by ($\frac{1}{3}\frac{1}{3}\tau$), with
$\tau$ incommensurate \cite{Oka08}: the limit of thick bilayers
seem to favor an incommensuration out-of-plane rather than
in-plane.

For the other extreme of thin bilayers, $R=$Y, the same type of
diffuse scattering along ($\frac{1}{3}\frac{1}{3}\ell$) is
observed at all $T$ (e.g.\ \cite{Horibe09}) for samples with
significant oxygen deficiency as in off-stoichiometric Lu or Yb
samples. However, the behavior of sufficiently stoichiometric
material is more complex. Abrupt changes of resistivity
accompanying the two magnetic transitions marked by arrows in
Fig.\ \ref{Inazumi} \cite{Tanaka82} suggest these are also
structural transitions, with monoclinic and finally triclinic
distortions as indicated by powder x-ray diffraction
\cite{Nakagawa79}.

Because weak superstructure peaks are difficult to observe in
powder diffraction, and sufficiently stoichiometric single
crystals were absent until recently, subsequent work on the CO
of stoichiometric YFe$_2$O$_4$ focused on electron-diffraction
(e.g.\ \cite{Horibe09,Ikeda03,Horibe10}). At low $T$, these
studies found clear superstructure spots with very different
propagation vectors, such as
($\frac{\overline{1}}{14}\frac{2}{7}\frac{1}{14}$) between $85$
and $130\,{\rm K}$, and consistent with triclinic symmetry. At
higher $T$ often a ($\frac{1}{4}\frac{1}{4}0$)-type
superstructure is observed, and sometimes
($\frac{1}{2}\frac{1}{2}0$)- or more ``conventional''
($\frac{1}{3}\frac{1}{3}\frac{3}{2}$)-types, partly with
different phases coexisting or relaxing from one to the other
as a function of time. The question of 3D CO in stoichiometric
YFe$_2$O$_4$ above the magnetic ordering temperature, in
particular at room temperature, is not clear, with either only
diffuse ($\frac{1}{3}\frac{1}{3}\ell$) lines reported
\cite{Ikeda03} or with coexistence of diffuse lines with broad
($\frac{1}{3}\frac{1}{3}\frac{n}{2}$) spots, the latter by
dark-field imaging assessed to correspond to 3D CO domains
correlated over some 30 bilayers \cite{Horibe09}.

These observations suggest a subtle competition of several CO
phases, that correspond to highly complex charge patterns, in
YFe$_2$O$_4$, reminiscent of some of the predictions based on
single-bilayer model Hamiltonians (see Sec.\ \ref{COtheory}).
To elucidate the strong difference in CO to LuFe$_2$O$_4$ will
require detailed modeling, but first the real-space CO giving
rise to the observed superstructure reflections  will have to
be solved, presumably by single-crystal x-ray diffraction on
sufficiently stoichiometric crystals \cite{MuellerDipl}.

\subsection{Fe-site substitution and intercalation}
\label{COYFO}

As mentioned in Sec.\ \ref{family}, possible changes in oxygen
stoichiometry hamper substitution studies on the Fe site.
Nevertheless, the expected quick depression of CO or charge
diffuse scattering seems to hold (see, e.g., \cite{Qin08}). For
intercalated materials, e.g.\ Lu$_2$Fe$_3$O$_7$, superstructure
reflections have been observed by electron diffraction,
interestingly up to $750\,{\rm K}$, and interpreted in terms of
both the bilayers and the intercalated single Fe-layers
exhibiting CO \cite{Qin09,Yang10,Qin10}. X-ray and neutron
diffraction on single crystals would certainly be desirable.

\begin{figure}[t]%
\sidecaption
\includegraphics*[width=0.57\linewidth]{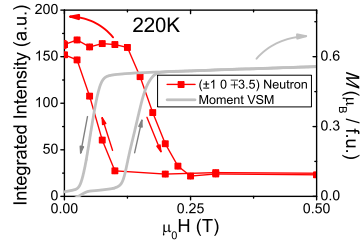}
\caption{%
  Magnetic field $H$ dependence of magnetization $M$ and the intensity of the ($\overline{1}02$)+($00\frac{3}{2}$) magnetic reflection at $220\, {\rm K}$
  in LuFe$_2$O$_4$. After \cite{DeGroot12}.}
\label{NDH}
\end{figure}

\begin{figure*}[bt]%
\sidecaption
\includegraphics*[width=0.73\linewidth]{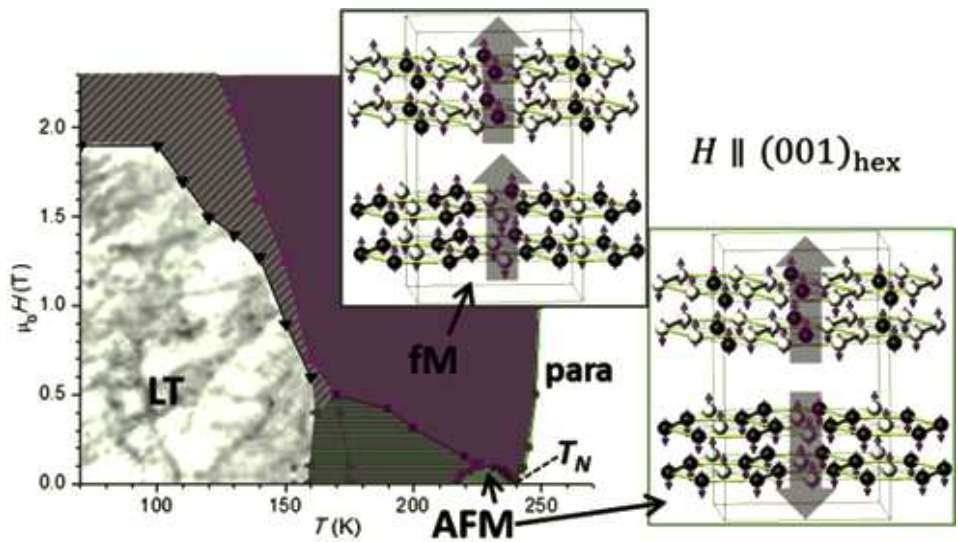}
\caption{%
  $H-T$ Magnetic phase diagram of LuFe$_2$O$_4$
  from data of \cite{DeGroot12,Xu08}. Shown are a low-$H$ antiferromagnetic (AFM) and a high-$H$ ferrimagnetic (fM) phase near the
  ordering temperature $T_N\sim 240\, {\rm K}$, as well as a lower-temperature phase (LT) with re-entrant disorder \cite{Christianson08}
  and a subtle structural modification \cite{Xu08}. Hysteretic regions where more than one phase can be stabilized are hatched.
  The ${\rm LT}\rightarrow{\rm fM}$ transition line ($\triangle$) reaches $7\,{\rm T}$ at $50\,{\rm K}$ (not shown) \cite{Xu08}. The
  spin structures of the AFM and fM phases \cite{DeGroot12} are also shown, with large semi-transparent arrows indicating the bilayer
  net magnetizations.}
\label{magphd}
\end{figure*}

\section{Magnetism and Spin-Charge coupling}

The $R$Fe$_2$O$_4$ family first attracted attention due to
intriguing magnetic properties of Ising spins (pointing
perpendicular to the layers) on a strongly frustrated lattice,
and the magnetic properties depend strongly on
oxygen-stoichiometry (c.f.\ Sec.\ \ref{family}).

Off-stoichiometric samples typically show significantly higher
magnetization when cooled in a magnetic field than when cooled
in zero field (c.f.\ Figs.\ \ref{Inazumi} and
\ref{LFOmagnChar}a), with a remanent magnetic moment after
cooling in strong magnetic fields between $1.5$ (Y) and
$2.8\,\mu_B/{\rm f.u.}$ (Lu) \cite{Iida89}.
Frequency-dependence in ac magnetic susceptibility is observed
below about $220-250\,{\rm K}$, with detailed analyses
suggesting either spin-glass- \cite{Wang09b} or
cluster-glass-like \cite{Wu08,Phan10,Sun12} freezing, with
large clusters of size $\sim\! 100\,{\rm nm}$ in-plane directly
observed by magnetic-force-microscopy in one example
\cite{Wu08}. Neutron diffraction studies suggest 2D magnetic
correlations, with diffuse magnetic scattering typically
observed along the ($\frac{1}{3}\frac{1}{3}\ell$) line
\cite{Funahashi84,Iida93} similar to the CO in
non-stoichiometric samples (c.f.\ Sec.\ \ref{COsec}). This
suggests that mainly the correlation between different bilayers
is lost. In less off-stoichiometric samples, a tendency towards
a development of peaks along this line has been observed
\cite{Funahashi84}, but so far few neutron diffraction studies
of stoichiometric crystals has been performed, and to date only
the magnetic structure of LuFe$_2$O$_4$ is known, which will
therefore be the focus in the following.

\subsection{3D magnetic correlations in LuFe$_2$O$_4$}
\label{3Dmag}

Sharp magnetic Bragg-reflections have been observed after
cooling below $T_N\sim 240\,{\rm K}$ at
($\frac{1}{3},\frac{1}{3},$integer) and
($\frac{1}{3},\frac{1}{3},$half-integer) positions in several
single crystals of LuFe$_2$O$_4$ by neutron diffraction
\cite{DeGroot12,Christianson08,Wen09,Wen10,Mulders11} and soft
resonant x-ray diffraction \cite{DeGroot12}. Magnetic
correlations between different bilayers have also been deduced
from powder neutron diffraction \cite{Wu08,Bourgeois12b}. An
example of ($\frac{1}{3}\frac{1}{3}\ell$) scans at different
temperatures is shown in Fig.\ \ref{XRDND}b): the weak
reflections at ($\frac{1}{3},\frac{1}{3},$half-integer) at
$280\,{\rm K}$ (thick olive) are due to CO, the much larger
intensity mainly at ($\frac{1}{3},\frac{1}{3},$integer) at
$220\,{\rm K}$ is magnetic in origin, as shown by polarized
neutron diffraction (see, e.g.\ Fig.\ 5 in \cite{DeGroot12}),
with an out-of-plane correlation length of $15\,{\rm nm}$. At
$130\,{\rm K}$ (red) a large diffuse background appears
together with peak-broadening, suggesting a re-entrant magnetic
disordering \cite{Christianson08,Wen10}. This low-temperature
state (LT phase) will be discussed in Sec.\ \ref{LT}.

In contrast to stoichiometric YFe$_2$O$_4$, where the absence
of any magnetic-field induced phase transitions is reported
\cite{Inazumi81,Nakagawa79}, magnetization-measurements on
stoichiometric LuFe$_2$O$_4$ reveal a complex magnetic field
($H\| (001)_{\rm hex}$)-temperature ($T$) phase diagram
\cite{DeGroot12,Xu08}, shown in Fig.\ \ref{magphd}. Apart from
the already mentioned LT phase with reentrant magnetic
disorder, and the paramagnetic state, two further phases are
present. Isothermal magnetization $M(H)$ loops (e.g.\ Fig.\
\ref{NDH} grey line) are consistent with a hysteretic
metamagnetic transition between a low-$H$ phase that is
antiferromagnetic (AFM) and a high-$H$ phase that is
ferrimagnetic (fM), with a net magnetic moment at low
temperatures (the fM phase remains stable down to $T=0$ when
cooled in $\mu_0 H>2\,{\rm T}$) of $\sim 3\mu_B/{\rm f.u.}$,
suggesting a $\uparrow\uparrow\downarrow$-like arrangement of
Fe spins. Partially strong hysteresis of the transition between
AFM and fM phases (hatched in Fig.\ \ref{magphd}), also present
between LT and either fM or AFM, is one of the unusual features
of the phase diagram. Magnetic Bragg reflections are observed
for both the AFM  and the fM phase at
($\frac{1}{3}\frac{1}{3}\ell$), with an intensity shift
surprisingly from $\ell$ integer to half-integer reflections
going from AFM to fM \cite{DeGroot12,Wen09}. As expected from
the presence of a net moment, in the fM phase magnetic
intensity is also observed on structural reflections. In
contrast, magnetic intensity at ($00\frac{3}{2}$)-type
reflections is a feature of the AFM phase (see Fig.\
\ref{NDH}).

\subsection{Competing spinstructures}
\label{spinstr}

Given the correspondence of Ising spins $\uparrow$ or
$\downarrow$ with valence states $2+$ and $3+$, both being
binary orders, the most straightforward way to analyze
spinstructures is to neglect the different spin quantum numbers
for the different valence states of the Fe ions and proceed
from the hexagonal cell analogous to the CO. As the same
propagation vectors are involved as considered for the CO in
Sec.\ \ref{LRCO}, there will again be three domains and the
same four possible orderings shown in Fig.\ \ref{COs} also then
apply to the spin order (SO), with black and white coloring now
indicating spins $\uparrow$ and $\downarrow$. If the weaker
($\frac{1}{3},\frac{1}{3},$half-integer) magnetic reflections
are considered to be due to a decoration by the CO, then the
proper SO should be the ordering shown in Fig.\ \ref{COs}
bottom left, because for a $\Gamma_1$ representation there
would be no intensity on ($\frac{1}{3}\frac{1}{3}0$), which is
the strongest magnetic reflection (Fig.\ \ref{XRDND}b). This
ferrimagnetic SO was found to fit well all
($\frac{1}{3},\frac{1}{3},$integer)-type reflections and was
thus initially proposed to apply \cite{Christianson08}.

However, the later found \cite{DeGroot12} magnetic intensity at
$(\overline{1}0\frac{7}{2})$ (Fig.\ \ref{NDH}) is inconsistent
with this SO, showing that symmetry-analysis based on the
hexagonal cell cannot describe the SO of LuFe$_2$O$_4$. In
retrospective this is not surprising, because the SO takes
place at $T_{\rm N} < T_{\rm CO}$ and if any spin-charge
coupling is present (c.f.\ Sec.\ \ref{SCC}) the proper cell on
which SO is to be considered is the CO supercell. Because the
magnetic cell is equal to the CO cell, symmetry analysis is
then based on a ($000$) propagation, and with four independent
sets of Fe sites present in this cell, and four irreducible
representations for each site, many SO have to be considered.
According to Landau-theory, all sites should order according to
the same irreducible representation if the ordering occurs in a
second-order phase transition at a single temperature. This
condition yields $40$ possible SO, which however were all found
to be inconsistent with the experimental data \cite{DeGroot12}.
In a brute-force approach all $3^{12}=531'441$ SO of Ising
spins ($\uparrow,\downarrow,0$, allowing for some Fe sites
remaining disordered) were compared with the data for both AFM
and fM phases, which identified the SO shown in Fig.\
\ref{magphd} as the correct ones, verified by refinement
\cite{DeGroot12}. For these SO, all four sets of Fe sites order
according to an irreducible representation, but with different
ones for different sets, implying that the SO does not occur
via a single second-order phase transition. Indeed, given that
both AFM and fM, separated by a first-order transition, are
present near $T_{\rm N}$ and $H=0$, a simple second-order
transition seems unlikely.

AFM and fM SO are closely related. In both cases, each bilayer
(and each individual layer) has a $2:1$
($\uparrow\uparrow\downarrow$ or
$\downarrow\downarrow\uparrow$) ferrimagnetic arrangement of
the Fe spins, with the bilayer net moment indicated by thick
semi-transparent arrows in Fig.\ \ref{magphd}. For the
ferrimagnetic fM phase, the net moments of the two bilayers of
the cell are both $\uparrow$, yielding a net moment, as
observed. For the AFM phase, the spins in one of the two
bilayers are exactly as in the fM phase, and for the other
bilayer they are all reversed, yielding the antiferromagnetic
behavior indicated by magnetization measurements. Intriguingly,
as can be seen in Fig.\ \ref{magphd}, the AFM-fM transition
extrapolates to $H\approx 0$ for $T\rightarrow T_{\rm N}$,
implying the near-degeneracy of these two SO phases at $T_{\rm
N}$ in zero field, a hallmark of geometrical frustration. A
competition of AFM and fM SO has also been observed in a study
on polycrystalline samples \cite{Bourgeois12b}. Because AFM and
fM SO are distinct only by the flip of all spins in one of the
bilayers, this suggests that the spin correlations within the
bilayers are much stronger than the inter-bilayer correlations.
In analogy to the CO (Sec.\ \ref{SRCO}), a random stacking of
the net moments of still relatively well spin-ordered bilayers
can then be expected above $T_{\rm N}$, with diffuse magnetic
scattering along ($\frac{1}{3}\frac{1}{3}\ell$). This has
indeed been observed (Fig.\ \ref{XRDND}c), up to room
temperature, and deviations of the magnetic susceptibility from
Curie-Weiss behavior suggest that spin-fluctuations correlated
within each bilayer remain relevant beyond even $T_{\rm CO}$
\cite{DeGroot12}. Note that these correlations are dynamic
(faster than $\sim$MHz) rather than static, because no
indications of magnetic order are visible in M{\"o}ssbauer
spectra at $260\,{\rm K}$ \cite{Xu08}.

\begin{figure}[t]
\includegraphics*[width=0.92\linewidth]{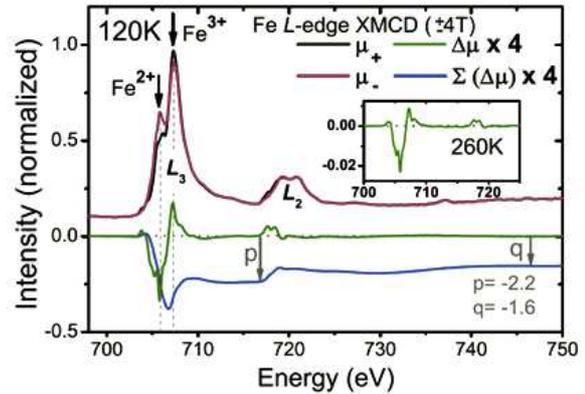}
\caption{%
  X-ray absorption spectra XAS with right ($\mu_+$) and left ($\mu_-$) circular polarized photons and X-ray
  Magnetic Circular Dichroism XMCD $\Delta\mu=\mu_+ - \mu_-$ of LuFe$_2$O$_4$ across the Fe $L_{2/3}$ edges
  in $4\, {\rm T}$ (perpendicular to the layers) at $120\, {\rm K}$ and   $260\, {\rm K}$ (inset). Also shown is the integration of the XMCD, with
  the sum rules yielding an orbital magnetic moment of $\sim\! 0.7\, {\mu_B /{\rm f.u.}}$. From \cite{DeGroot12b}, $\copyright \ $ 2012
  American Physical Society.}
\label{XMCD}
\end{figure}

\subsection{Spin-charge coupling}
\label{SCC}

\paragraph{Experimental indications}

Experimental indications of a spin-charge coupling by anomalies
in the $T$-dependence of the intensity or incommensuration of
the CO superstructure reflections at $T_N$ have been observed
\cite{Angst08,Bartkowiak12}, but they are weak. However, this
might be expected if the SO within individual bilayers is
formed already above $T_N$. The SO refinements \cite{DeGroot12}
yielded some hints of a strict assignment of SO and CO, but the
different magnetic moments of Fe$^{2+}$ and Fe$^{3+}$ ions
could not be reliably determined. However, spectroscopic
techniques performed on a variety of samples provide strong
evidence that such spin-charge coupling is a common feature at
least in YbFe$_2$O$_4$ and LuFe$_2$O$_4$
\cite{DeGroot12b,Ko09,Kuepper09,Tanaka89,Nakamura98}. In the
first studies, polarized M{\"o}ssbauer spectroscopy was applied
to single crystals of YbFe$_2$O$_4$ \cite{Tanaka89} and
LuFe$_2$O$_4$ \cite{Nakamura98} under varying $H$, it was
concluded that all of the Fe$^{2+}$ spins as well as
$\frac{1}{3}$ of the Fe$^{3+}$ spins were aligned $\|H$, with
the remaining $\frac{2}{3}$ of the Fe$^{3+}$ spins aligned
antiparallel to $H$.

Later studies by x-ray magnetic circular dichroism (XMCD)
confirmed this arrangement for LuFe$_2$O$_4$
\cite{DeGroot12b,Ko09,Kuepper09}. Fig.\ \ref{XMCD} shows the
absorption spectra and XMCD measured (in the fM-phase) on a
crystal of the same type as used to establish the magnetic
phase diagram and SO (c.f.\ Fig.\ \ref{magphd}). The larger
negative peak of the XMCD signal at the Fe$^{2+}$ $L_3$ edge
and the smaller positive peak at the Fe$^{3+}$ $L_3$ edge
directly implies a larger Fe$^{2+}$ net moment in magnetic
field-direction, and a smaller Fe$^{3+}$ net moment opposite to
this, which given the known magnetic cell-size and saturation
magnetization leaves the above spin configuration as the only
possible one. From the XMCD sum rules, a sizeable orbital
magnetic moment of $\sim\! 0.7-0.8\, {\mu_B /{\rm Fe}^{2+}}$
ion can be concluded \cite{DeGroot12,Ko09,Kuepper09}, rendering
the Fe$^{2+}$ total magnetic moment to $\sim\! 4.75\, \mu_B$,
quite close to the spin-only Fe$^{3+}$ moment of $5\, \mu_B$.
The resulting saturation moment of $4.75-\frac{1}{3}\cdot5 \sim
3.08\, {\mu_B /{\rm f.u.}}$ is in agreement with magnetization
data.

In Fig.\ \ref{magphd}, both spin and charge structures are
shown, and it can be readily seen that for the fM-phase the
alignment of Fe$^{2+}$ and Fe$^{3+}$ moments according to XMCD
and M{\"o}ssbauer spectroscopy is indeed observed, i.e. the
previously discussed CO and SO are consistent with strict
spin-charge coupling as deduced spectroscopically. Moreover,
the SO shown in Fig.\ \ref{magphd} combined with the
spectroscopic results is incompatible with any of the CO shown
in Fig.\ \ref{COs} except the charged bilayer stacking deduced
in Sec.\ \ref{LRCO}, providing additional support for this
model of CO \cite{DeGroot12b}.

The inset of Fig.\ \ref{XMCD} shows a XMCD spectrum measured in
$4\,{\rm T}$ above $T_{\rm N}$, at $260\,{\rm K}$. Though much
reduced in amplitude, the shape is still the same as in the
fM-phase below $T_{\rm N}$. This is consistent with the model
of randomly-stacked but otherwise still well spin-ordered
bilayers in the paramagnetic state: Polarized by a sufficiently
strong magnetic field, the random-stacking of 50\% $\uparrow$
and 50\% $\downarrow$ bilayer net moments will be broken, with
the overall spin arrangement tending towards the fM SO. It also
shows that the spin-charge coupling is very robust, not
requiring long-range order. The magnetic correlations could
thus be important not only for the magnetic transition at
$T_{\rm N}$, but also for the charge ordering at $T_{\rm CO}$.

\begin{figure}[t]%
\includegraphics*[width=0.9\linewidth]{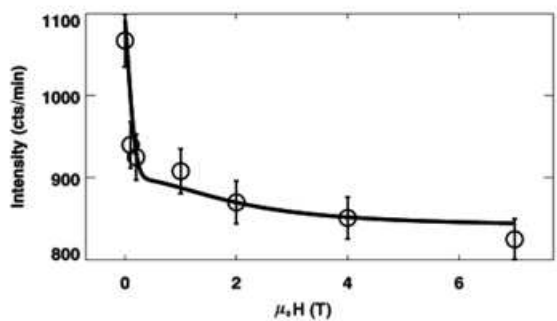}
\caption{%
  Intensity of the ($\frac{2}{3}\frac{2}{3}\frac{7}{2}$) CO superstructure peak measured by
  neutron diffraction at $300\, {\rm K}$, after cooling in magnetic fields of different strengths. From \cite{Wen09},
  $\copyright \ $ 2009 American Physical Society.}
\label{HdepCO}
\end{figure}

Interestingly, a magnetic-field influence on the CO has been
observed for LuFe$_2$O$_4$ \cite{Wen09}: Fig.\ \ref{HdepCO}
shows the intensity of the
($\frac{2}{3}\frac{2}{3}\frac{7}{2}$) CO superstructure
reflection, measured at $300\,{\rm K}$, after cooling the
crystal in different magnetic fields from $350\,{\rm K}$. A
significant suppression of $\sim\! 15\%$ already by rather
small fields is visible. This is surprising, given that at the
AFM-fM metamagnetic transition no change of the CO was observed
\cite{DeGroot12b}. However, in contrast to these measurements
with $H\|c$, in \cite{Wen09} the field was applied in
$[1\overline{1}0]$ direction. This in-plane $H$ breaks the
three-fold rotation symmetry of the crystal, and likely affects
the relative stability of the three magnetic domains (based, as
the CO domains, on the three monoclinic cells shown in Fig.\
\ref{compstrucfig}a). The strict spin-charge coupling implies
the correspondence of magnetic and CO domains, and therefore an
in-plane $H$ should affect the domain population. The
($\frac{2}{3}\frac{2}{3}\frac{7}{2}$) reflection studied in
\cite{Wen09} belongs to one of the CO domains, its intensity
suppression by $H$ suggests that population of other CO domains
is favored, which could be tested by simultaneously tracking
different CO reflections belonging to all domains.

\paragraph{Theoretical considerations}

The single-bilayer lattice gas models
\cite{Nagano07,Naka08,Nasu08,Watanabe09,Watanabe10} also
consider different SO and spin-charge coupling, finding that
the spin-correlations tend to stabilize CO with
($\frac{1}{3}\frac{1}{3}\ell$)-type propagation compared to
alternatives (c.f.\ Sec.\ \ref{COtheory}). However, because
these models do not consider the charged bilayers indicated by
experiment, the focus below is on more phenomenological
considerations and the DFT calculation \cite{Xiang09}.

Because the inter-bilayer spin-coupling is very weak (c.f.\
Sec.\ \ref{spinstr}), it is a good first approximation to
consider each bilayer, the one with Fe$^{2+}$ majority and the
one with Fe$^{3+}$ majority, individually. Various magnetic
superexchange parameters between Fe ions of the same layer and
between the two layers have then to be examined, which may be
expected to strongly depend on the valence state of the
respective Fe ions. An experimental determination of exchange
parameters could be done by inelastic neutron scattering, which
so far have been reported only in a limited range on a crystal
\cite{LawrencePhD}, and on a polycrystalline sample, with the
observation of a spin-gap $\sim\! 7\,{\rm meV}$ confirming the
Ising nature of spins \cite{Bourgeois12b}.

The exchange parameters were estimated theoretically by mapping
the energy differences between differently ordered states
obtained by DFT to the corresponding energy differences of an
Ising Hamiltonian \cite{Xiang09}. This study considered an
individual bilayer, however, one assumed to exhibit the
ferroelectric CO (Fig.\ \ref{COs} top left). The by far largest
interaction was found to be an antiferromagnetic superexchange
between Fe$^{3+}$ ions of the same Fe$^{3+}$ majority layer.
From this, an in-plane antiferromagnetic arrangement of
Fe$^{3+}$ moments is expected, which is indeed observed
experimentally (see Fig.\ \ref{magphd}). A corresponding
calculation taking into account the charged bilayer CO would be
highly desirable: as mentioned already in Sec.\ \ref{COtheory},
the large energy-differences between different spin-states
found in \cite{Xiang09} imply that the CO and SO have to be
calculated together, in a cell of sufficient size, which makes
the calculations challenging.

\subsection{Off-stoichiometric samples and other rare earths}

The agreement between XMCD experiments conducted on samples
with \cite{DeGroot12b} and without \cite{Ko09} long-range CO
indicates that the strict spin-charge coupling with the
Fe$^{2+}\uparrow\uparrow\uparrow\,
/$Fe$^{3+}\uparrow\downarrow\downarrow$ arrangement is a robust
and generic feature of LuFe$_2$O$_4$. Because the superexchange
interactions leading to spin-charge coupling and SO depend on
the underlying CO, it seems very likely that in all samples
individual bilayers exhibit both CO and SO as discussed in
Secs.\ \ref{LRCO}, \ref{SRCO}, and \ref{spinstr}, with only the
occurrence of long-range 3D order with correlations between
different bilayers being affected by stoichiometry. The generic
observation of at least diffuse charge (e.g.\ \cite{Wu08}) and
diffuse magnetic (e.g.\ \cite{Iida93}) scattering at the
($\frac{1}{3}\frac{1}{3}\ell$) line supports this assessment.
Because the charge dynamics slows down upon cooling regardless
of stoichiometry \cite{Tanaka84}, a glassy freezing of the
essentially still charge-ordered bilayers can be expected to
replace the long-range CO when the latter is prevented.
Similarly, a glassy freezing of SO bilayers can be expected to
replace the long-range AFM phase upon cooling, and such a
cluster-glass-like behavior involving frequency-dependence in
ac susceptibility has indeed been found for some samples, e.g.\
in \cite{Phan10}. On the other hand, when a magnetic field
forces the net moments of all bilayers to align, a stable
fM-like phase with the same saturation moment should be
recovered, as is indeed suggested by all macroscopic
measurements (see e.g.\ \cite{Wang09b,Phan10} for corresponding
phase diagrams) and by electron holography \cite{Maruyama12}.
Note, however, that because of the underlying random stacking
of Fe$^{2+}$ and Fe$^{3+}$ majority bilayers, this fM phase is
not expected to be 3D ordered, but should rather exhibit
diffuse magnetic scattering along
($\frac{1}{3}\frac{1}{3}\ell$), which could be checked by
in-field neutron diffraction on less-stoichiometric crystals.
Thus, based on the above, for off-stoichiometric samples,
exactly the ``spin-glass-like 2D-ferrimagnetic order'' proposed
early on \cite{Iida93} is expected to occur.

The above considerations should hold also for YbFe$_2$O$_4$,
which has a similar $R$ ion size. Indeed, not only the same
diffuse scattering at ($\frac{1}{3}\frac{1}{3}\ell$)  has been
observed \cite{Hearmon12} as on non-stoichiometric
LuFe$_2$O$_4$, but the same spin-charge coupling is also
deduced \cite{Tanaka89}. For $R$=In with considerably smaller
ion size, magnetism and spin-charge coupling appear to be
rather different as indicated both by a much smaller saturation
moment of $1.5\,\mu_B/{\rm f.u.}$ and dissimilar M{\"o}ssbauer
spectra \cite{Oka08}. An XMCD study on InFe$_2$O$_4$ might be
helpful to elucidate spin-charge coupling in this material. For
$R$ with considerably larger ionic radii than Lu or Yb, there
are also clear differences. For example, in stoichiometric
YFe$_2$O$_4$ there is no fM phase, and given the wildly varying
CO observed at low $T$ (see Sec.\ \ref{COYFO}) a similar
spin-charge coupling as in LuFe$_2$O$_4$ is rather doubtful,
despite similarities of diffuse scattering observed in
off-stoichiometric crystals. In agreement with substantial
changes upon increasing the rare earth ionic size, an XMCD
study on LuFe$_2$O$_4$ with Lu partially substituted by the
larger Y or Er ions \cite{Noh10}, found marked changes in the
spectra. Whereas the negative peak at the Fe$^{2+}$ edge
remained largely unaffected by the substitution, suggesting
that ``all Fe$^{2+}$ spins aligned with $H$'' still holds, the
positive peak at the Fe$^{3+}$ edge was completely suppressed
for $\sim 50\%$ substitution, suggesting an antiferromagnetic
order of Fe$^{3+}$ spins.

\section{The ground state: orbital order?}
\label{OO}

\begin{figure}[bt]%
\includegraphics*[width=0.7\linewidth]{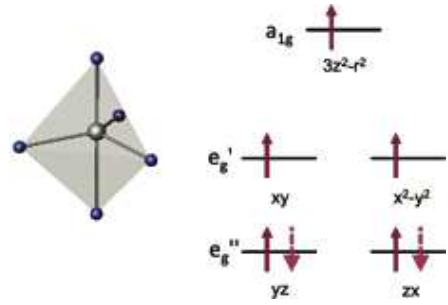}
\caption{%
  Sketch of the energy-splitting of the Fe $3d$ orbitals (right)
due to the trigonal-bipyramidal coordination (left). For
Fe$^{3+}$, each orbital is singly occupied, the extra-electron
of Fe$^{2+}$ can go to either of two $e_g ''$ orbitals.}
\label{xtalfield}
\end{figure}

\subsection{Orbital order, orbital magnetic momentum, and magnetic anisotropy}
\label{orbmag}

Concomitant with CO and SO, orbital order (OO) typically occurs
in other transition metal oxides, strongly influencing the
magnetism as well \cite{Tokura00}. The possible presence of OO
has been considered for LuFe$_2$O$_4$ theoretically
\cite{Xiang07,Nagano07,Nasu08}, for example the DFT
calculations \cite{Xiang07} suggest a clear OO. However, the
discrepancy of the predicted CO with the experimentally deduced
one suggests that the predictions of OO also have to be taken
with care.

The natural starting point for the consideration of OO is the
trigonal bipyramidal crystal-field level diagram shown in Fig.\
\ref{xtalfield}. For Fe$^{3+}$, all five orbitals are singly
occupied by a spin $\uparrow$ electron, and there is no orbital
degree of freedom. However, the extra $\downarrow$-electron at
Fe$^{2+}$ ions can go to an orbital of the doubly-degenerate
lowest level. It should be mentioned that $e_g'$ and $e_g''$
levels are quite close in energy in LuFe$_2$O$_4$: while
band-structure calculations suggested that $e_g'$ is the lowest
level \cite{Xiang07}, polarized x-ray absorption spectroscopy
at the O $K-$edge has later shown that the $e_g''$ level is
actually lower \cite{Ko09}.

OO would then imply a preferential occupation of either the
$yz$ or the $zx$ orbital $-$ or a given real linear combination
of these $-$ for each site. An alternative to this is the
preferential occupation of a {\em complex} linear combination
that could carry an orbital angular momentum $\mathbf{L}$
($\left \langle \psi \right | \mathbf{L} \left | \psi \right
\rangle =0$ for $\psi$ real) and thus an orbital magnetic
moment. From a linear combination of $yz$ and $zx$ orbitals, no
eigenstates of $L_x$ or $L_y$ can be obtained, but
$\sqrt{\frac{1}{2}} \left ( \mp zx - i\cdot yz \right )$ is an
eigenstate of $L_z$ with eigenvalue $m=\pm 1$, allowing an
orbital magnetic moment of up to $\pm 1\mu_B/{\rm Fe}^{2+}$
ion. The fact that an orbital magnetic moment $\sim\! 0.7-0.8\,
{\mu_B /{\rm Fe}^{2+}}$ {\em is} observed (see Sec.\ \ref{SCC})
implies that in the fM state (and at low $T$ for samples
without long-range SO) OO cannot occur. Consistent with this,
no additional reflections have ever been found, and no
anisotropies in resonant diffraction suggestive of ferro-type
OO have been found \cite{Mulders09}. For samples with
long-range SO, this conclusion can also be extended to the AFM
and paramagnetic phases (c.f.\ Fig.\ \ref{magphd}), because
high-resolution x-ray diffraction indicates no structural
changes that necessarily would accompany an OO-transition at
the fM-AFM or AFM-paramagnetic transitions \cite{DeGroot12b}
(though for the latter small intensity anomalies are suggested
in \cite{Angst08,Bartkowiak12}).

From the presence of an orbital magnetic moment necessarily
perpendicular to the layers, the large magnetic anisotropy
making this a model Ising-system follows immediately, because
spin-orbit coupling aligns the directions of spin- and orbital
magnetic moments \cite{Ko09}. The large magnetic anisotropy is
to a good part responsible for the giant coercivity of
LuFe$_2$O$_4$, with coercive fields reaching around $10\,{\rm
T}$ at low $T$ \cite{Iida87,Wu08}, and to
``exchange-bias''-like properties \cite{Yoshii12}.

\subsection{The low-temperature phase}
\label{LT}

The only phase in the $H-T$ phase diagram (Fig.\ \ref{magphd})
consistent with OO is the LT-phase, because both ${\rm
AFM}\leftrightarrow{\rm LT}$ and ${\rm fM}\leftrightarrow{\rm
LT}$ transitions are also structural transitions as indicated
by Bragg-peak splitting \cite{Xu08}. As discussed in Sec.\
\ref{SRCO}, the transition to this phase 
is accompanied by the complete freezing of residual charge
dynamics, which is consistent with an OO scenario, given that
OO often accompanies metal-insulator transitions
\cite{Tokura00}. Due to the impact of OO on the magnetic
exchange interactions, a change of the magnetic structure upon
entering the LT phase would then be expected, consistent with
the observed re-entrant disorder (Fig.\ \ref{XRDND}b).

These indications that OO might be the driving force behind the
transition into the LT phase are, however, at most tentative.
As discussed in Sec.\ \ref{orbmag}, a full OO would imply the
quenching of the orbital magnetic moment and the removal of the
strong tendency of spins to be aligned perpendicular to the
layers. The relatively good description of the
$\ell-$dependence of the diffuse component of magnetic
scattering at $130\,{\rm K}$ assuming spin direction
perpendicular to the layers \cite{Christianson08} is therefore
an argument against an OO scenario.

Structurally, despite of the Bragg-peak splitting, there is no
basic change of the CO across $T_{\rm LT}$: the interbilayer
correlations of CO seem to improve slightly \cite{Angst08}, but
the cell size (no extra reflections), $C2/m$ symmetry, and the
valence states of the Fe ions estimated by bond-valence-sum are
unaffected within error bars \cite{DeGroot12b}. Although small
changes in lattice parameters have been reported, the changes
of refined atom positions are much too subtle to conclude about
a possible OO \cite{DeGroot12b}. Further structural refinement
studies across $T_{\rm LT}$ are certainly desirable. A further
possibility to detect OO is to assess asymmetries in resonant
x-ray diffraction, ideally at the Fe $L$ edges at the
($\tau\tau\frac{3}{2}$) reflections \cite{DeGrootPhD,BlandPhD},
although the analysis is complicated by the possible emergence
of magnetic intensity at these positions in the LT phase
\cite{Christianson08,Wen10}.

The nature of the LT phase, and whether it is connected with OO
therefore is an open question left for future work. Its
experimental investigation is also hampered by its very fragile
nature, as it is present in some samples showing long-range CO
and SO \cite{DeGroot12b,DeGroot12,Christianson08,Wen10,Xu08}
(indications are also visible in $M(T)$ on the polycrystalline
samples studied in \cite{Iida86,Bourgeois12b}), but not in
others \cite{Yamada00,Wen09}. Due to this, it is also not
completely clear whether the LT phase represents the true
ground state of LuFe$_2$O$_4$, though it appears likely. If the
LT phase {\em is} the true ground state, it will make
theoretical investigations with zero-temperature methods, such
as DFT, difficult because of the presence of significant
magnetic disorder.

\section{Summary and outlook}
\label{sum}

In the following, I will summarize the main results reviewed
and {\em highlight} critical points for future investigations
on rare earth ferrites, concluding by answering the title
question and by looking at the larger picture of possible
CO-based ferroelectricity in other compounds.

For the most studied rare earth ferrite, LuFe$_2$O$_4$, a
consistent and rather comprehensive picture of strongly coupled
charge- and spinorder has emerged from experimental work.
Single-crystal x-ray and neutron diffraction revealed sharp
charge order (CO) superstructure peaks below $T_{\rm CO}\sim\!
320\, {\rm K}$ and spin order (SO) superstructure peaks below
$T_{\rm N}\sim\! 240\, {\rm K}$ in stoichiometric samples. Both
are unaffected by static electric fields or currents, except
for self-heating effects. The corresponding orders have been
determined by refinements. The CO involves charged, rather than
polar, bilayers. As revealed by M{\"o}ssbauer and optical
spectroscopies, substantial charge fluctuations occur even
below $T_{\rm CO}$, down to $T_{\rm LT}\sim\! 170\, {\rm K}$.
Between $T_{\rm LT}$ and $T_{\rm N}$ two related SO phases
compete, due to geometrical frustration. {\em Below $T_{\rm
LT}$}, a subtle structural distortion and re-entrant magnetic
disorder occur. Whether this is connected with {\em possible
orbital ordering is a major question left for future work}.
M{\"o}ssbauer spectroscopy and x-ray magnetic circular
dichroism demonstrate a strict coupling of SO and CO, present
also above $T_{\rm N}$ and likely even above $T_{\rm CO}$,
where diffuse scattering experiments indicate that within
individual bilayers SO and CO correlations are still strong.
The presence of coupled SO and CO correlations within
individual bilayers is a robust feature of the compound.
However, oxygen off-stoichiometry quickly destroys the
inter-bilayer spin- and charge-correlations, leading to
``glassy'' behavior.

While charge and spin order have been established, they are not
really understood: this would imply elucidating what drives
these orders, whereas currently no theoretical works have
considered the experimental charge or spin order. What is clear
is that the charge order cannot be driven entirely or even
predominantly by electrostatic repulsion, because the realized
pattern is not the one minimizing this repulsion. Lattice
effects as well as magnetic exchange appear to be relevant, but
{\em the determination of the main driving forces remains a
challenge principally for future theoretical work}, though {\em
experimental work on phonons and magnetic excitations} could
make an important contribution: this is {\em the most critical
issue for future investigations}.

{\em Rare earth ferrites other than LuFe$_2$O$_4$ have been
studied much less}, which is largely due to the absence of
single crystals that are stoichiometric enough to reveal 3D CO
and SO. With this hurdle being overcome \cite{MuellerDipl},
progress can be expected. CO and SO correlations leading to
diffuse scattering very similar as in off-stoichiometric
LuFe$_2$O$_4$ are commonly observed, and for YbFe$_2$O$_4$ also
M{\"o}ssbauer spectroscopy leads to the expectation of overall
behavior very similar to LuFe$_2$O$_4$. However, for rare
earths (or In) with larger deviations of ionic size from Lu,
experiments on high-quality polycrystalline samples indicate
that completely different CO and SO is established. For
example, electron diffraction on {\em YFe$_2$O$_4$}
demonstrates a number of {\em competing very complex CO phases
that have yet to be solved}. A partial alternative to
single-crystal work is to rely on experimental techniques not
requiring long-range CO and SO, such as XMCD, which has been
successfully used to study ion-size effects on spin-charge
coupling \cite{Noh10}. The {\em intercalated compounds
$R_2$Fe$_3$O$_7$} have been even less studied, but given
indications of charge order stable up to much higher
temperatures \cite{Yang10} {\em deserve more attention}. A
similar tuning as with intercalation may be possible with {\em
thin films} \cite{Zeng12,Rai12,Wang12b,Brooks12}, on which,
however, {\em 3D charge ordering has yet to be demonstrated}.

\paragraph{Ferroelectricity from charge ordering?}

While the coupled complex charge and spin orders on a
geometrically frustrated lattice are also of high intrinsic
interest, it is not the property that has attracted most
research to the rare earth ferrites. The title question
``Ferroelectricity from iron valence ordering in rare earth
ferrites?'' according to current experimental knowledge has to
be answered in the negative, at least for the ``prototype''
LuFe$_2$O$_4$. The charge order leading to the
($\frac{1}{3}\frac{1}{3}\ell$)-type reflections does not
involve polar bilayers, and according to recent dielectric
spectroscopy studies, the large remanent polarization deduced
from pyroelectric current measurements may likely be affected
by contact effects. {\em Future theoretical work leading to an
understanding of what drives the experimentally indicated
non-polar charge order would be important for a final
resolution and to provide hints for other families of
compounds}. In rare earth ferrites other than LuFe$_2$O$_4$ the
charge order can be different and has not been solved,
therefore in principle the possibility of ferroelectricity from
charge ordering remains within the family. However, with
dielectric behavior quite common among other $R$Fe$_2$O$_4$ and
even the most likely ferroelectric behavior in a doped
intercalated compound suggesting a polarization three orders of
magnitude weaker than what is expected for ferroelectricity
originating from charge ordering, there are no strong
indications for success.

Looking beyond rare earth ferrites, the findings reviewed here
provide some general lessons for CO-induced ferroelectricity.
First and most important, a non-negligible residual
conductivity is unavoidable in CO materials, particularly at
and above $T_{\rm CO}$. Apart from being detrimental with
regards to applications, this also renders the standard
macroscopic probes of ferroelectricity doubtful as pointed out
in \cite{Maglione08} and for rare earth ferrites discussed in
Sec.\ \ref{macroFE}. Therefore it is imperative to obtain also
microscopic proof of at least a polar crystal structure, better
yet of a polar crystal structure that can be switched by an
electric field $-$ though for such in-electric field
measurements residual conductivity and associated self-heating
effects are also problematic, see Sec.\ \ref{COinE}. Second, CO
materials have a propensity for variable stoichiometry, which
can make the elucidation of the CO crystal structure very
difficult. Therefore, paying due respect to sample quality is
crucial. Finally, in CO materials there is often a strong
coupling of charge with orbital and spin degrees of freedom
leading to complexity that hampers theoretical treatments. For
rare earth ferrites, theoretical work suggests and/or assumes
ferroelectric CO (see Sec.\ \ref{COtheory}), in contrast to the
experimental findings. For other materials, there are many more
examples of compounds proposed to exhibit ferroelectricity from
CO based on calculations than examples of experimental
indications of the same.

With the prototypical example of rare earth ferrites apparently
being a non-example, a valid ``proof-of-principle'' example of
a material exhibiting multiferroicity driven by charge ordering
has yet to be demonstrated experimentally. Among the few
possibilities in oxides with at least some experimental
indications most promising seems classical magnetite, which
{\em does} seem to have a charge-ordered crystal structure that
is polar \cite{Senn12}, although it appears to be of a
relaxor-type with switching only possible at low $T$ and high
frequency \cite{Schrettle11}. Apart from oxides,
low-dimensional molecular compounds (see e.g.\
\cite{Monceau12}) are increasingly being scrutinized for
multiferroicity originating from CO. A recently proposed
\cite{Lunkenheimer12}, but also disputed \cite{Sedlmeier12},
example is a metal-organic charge-transfer salt with apparent
small polarization but no demonstrated polar crystal structure.
Although the temperature scale in all of the above is far below
room temperature, an unambiguous proof of the mechanism would
be important, opening the route to search for such materials
closer to applicability.

\begin{acknowledgement}
I thank J.~de Groot, T.\ Mueller, R.~P.\ Hermann, and in
particular Th.\ Br{\"u}ckel, for useful comments on draft
versions of this article, T.\ M{\"u}ller also for digitizing
the data from \cite{Inazumi81} shown in Fig.\ \ref{Inazumi}. I
am very grateful to my students and to the many collaborators I
had while working in this field, in particular the coauthors of
Refs.\
\cite{DeGroot12b,Niermann12,Xu10,Phan09,DeGroot12,Angst08,Christianson08,Xu08,Phan10}.
Special thanks go to D.~Mandrus for first suggesting this field
to me, to M.~Subramanian for providing high-quality
LuFe$_2$O$_4$ powder samples very early-on, and to J.~de Groot
for his immense contributions during his PhD studies. Finally,
I would like to acknowledge financial support from the
initiative and networking fund of the Helmholtz Association by
funding the Helmholtz-University Young Investigator Group VH
NG-510 ``Complex Ordering Phenomena in Multifunctional
Oxides''.
\end{acknowledgement}

\newcommand{\noopsort}[1]{} \newcommand{\printfirst}[2]{#1}
  \newcommand{\singleletter}[1]{#1} \newcommand{\switchargs}[2]{#2#1}
\providecommand{\WileyBibTextsc}{}
\let\textsc\WileyBibTextsc
\providecommand{\othercit}{} \providecommand{\jr}[1]{#1}
\providecommand{\etal}{~et~al.}

\end{document}